\documentclass[graybox, natbib]{svmult}
\bibpunct{(}{)}{;}{a}{}{,} % suppress commas between author-nMames and year
\usepackage{xcolor} %for colored notes
\usepackage{bm}
\usepackage{helvet}           % selects Helvetica as sans-serif font
\usepackage{courier}          % selects Courier as typewriter font
\usepackage{type1cm}          % activate if the above 3 fonts are
\usepackage{makeidx}          % allows index generation
\usepackage{graphicx}         % standard LaTeX graphics tool
\usepackage{multicol}         % used for the two-column index
\usepackage[bottom]{footmisc} % places footnotes at page bottom
\usepackage[normalem]{ulem}	  % for strike-through of text with \sout{}
\usepackage{soul}             % for high-lighting of text
\usepackage{amsmath}
\usepackage{amssymb}
\usepackage{hyperref}         % for hyperlinks

\usepackage{bm}
\usepackage[labelfont={small,bf},textfont=small]{caption}
\usepackage{subcaption}
\usepackage{colortbl}
\usepackage[margin=10pt,font=small,labelfont=rm,labelsep=endash]{caption}

\def\smallgamma{{\scriptscriptstyle \gamma}}
\def\smallone{{\scriptscriptstyle 1}}
\def\smallE{{\scriptscriptstyle E}}
\def\smallQ{{\scriptscriptstyle Q}}
\def\smallZ{{\scriptscriptstyle Z}}
\def\smallXC{{\scriptscriptstyle XC}}
\def\smallDFT{{\scriptscriptstyle DFT}}
\def\smallGGA{{\scriptscriptstyle GGA}}
\def\smallH{{\scriptscriptstyle H}}
\def\smallKS{{\scriptscriptstyle KS}}

\def\eqref#1{(\ref{#1})}
\def\angstrom{{\mbox{\AA}}}
\newcommand{\un}[1]{\,\mathrm{#1}}
\def\rGamma{{\mathrm\Gamma}}
\def\rDelta{{\mathrm\Delta}}
\def\rLambda{{\mathrm\Lambda}}
\def\rOmega{{\mathrm\Omega}}

\makeindex             % used for the subject index
\begin{document}

\title*{Heat transport in insulators \\ from ab initio Green-Kubo theory}
\author{Stefano Baroni, Riccardo Bertossa, Loris Ercole, Federico Grasselli, and Aris Marcolongo}
\authorrunning{S. Baroni \emph{et al.}}
\titlerunning{Ab initio Green-Kubo theory}
\institute{Stefano Baroni \at SISSA -- Scuola Internazionale Superiore di Studi Avanzati and \\ CNR Istituto Officina dei Materiali \\ 34136 Trieste, Italy \email{stefano.baroni@sissa.it}
\and Riccardo Bertossa, Loris Ercole, and Federico Grasselli \at SISSA
\and Aris Marcolongo \at
Cognitive Computing and Computational Sciences Department, IBM Research \\ 8803 Z\"{u}rich, Switzerland \\
THEOS-MARVEL, \'Ecole Polytechnique F\'ed\'erale de Lausanne \\ 1015 Lausanne, Switzerland
}

\maketitle

%%%%%%%%%%%%%%%%%%%%%%%%%%%%%%%%%%%%%%%%%%%%%%%%%%%%%%%%%%%%%%%%%%%%%%%%%%%%%%%%%%%%
%%%%%%%%%%%%%%%%%%%%%%%%%%%%%%%%%%%%%%%%%%%%%%%%%%%%%%%%%%%%%%%%%%%%%%%%%%%%%%%%%%%%
Heat conduction in insulators is determined by the dynamics of the atomic nuclei, the electrons following adiabatically in their ground state: a regime that we will refer to as atomic or \emph{adiabatic heat transport}. When a quasi-particle picture of the heat carriers (phonons) is adequate, a kinetic approach to heat conduction based on the Boltzmann transport equation \citep{Peierls1929} has been demonstrated \citep{Broido:2007iu} and successfully applied to crystalline insulators \citep{Zhou2016}, leveraging phonon frequencies and lifetimes from density-functional perturbation theory \citep{Baroni1987a,Giannozzi1991,Debernardi1995,Baroni2001}. As the temperature increases, anharmonic effects eventually make the quasi-particle picture break down well below melting \citep{Turney:2009bb}, while the Boltzmann transport equation simply does not apply to glasses and liquids, where phonon dispersions are not even defined.

The simulation of thermal transport in glasses, liquids, and high-tem\-per\-a\-ture crystalline solids thus requires the full deployment of the statistical mechanics of hydrodynamic fluctuations \citep{Kadanoff1963}, started in the thirties by Lars Onsager \citep{Onsager1931a,Onsager1931b} and culminated in the fifties with the Green-Kubo (GK) theory of linear response \citep{Green1952,Green1954,Kubo1957a,Kubo1957b}. According to this theory, the thermal conductivity, $\kappa$, can be expressed in terms of time correlations of the heat current, $J$, as:
\begin{equation}
\kappa\propto\int_{0}^{\infty}\!\langle{J}(t){J}(0)\rangle\, dt,\label{eq:GK}
\end{equation}
where the brackets indicate ensemble averages over trajectories, which are accessible in principle to molecular dynamics (MD) simulations. In spite of the beauty, rigor, and broad scope of the GK theory, its adoption in the \emph{ab initio} simulation of heat transport has stumbled at two seemingly insurmountable hurdles: first and foremost, the heat current entering Eq.~\eqref{eq:GK} is ill-defined at the atomic scale, thus allegedly thwarting its implementation from first principles; second but not less important, experience in classical simulations, where a definition of the heat current is readily available in terms of suitably defined atomic energies \citep{Irving1950} indicates that taming its statistical fluctuations requires so long MD simulations, as to be unaffordable using \emph{ab initio} (AI) MD \citep{Car1985,Marx2009}. As a matter of fact, no AIMD simulations of adiabatic heat transport based on GK theory have appeared until the spurious nature of these hurdles was recently revealed by \cite{Marcolongo2016} and \cite{Ercole2016,Ercole2017}.

The first difficulty was overcome by revealing a general \emph{gauge invariance} principle for thermal transport, stemming from the hydrodynamic nature of energy fluctuations (see below), which makes transport coefficients independent of the microscopic expression of the energy density and current that are to a large extent ill defined \citep{Marcolongo2016,Ercole2016}. Building on this principle, an expression for the energy current was derived from density-functional theory \citep{Hohenberg1964,Kohn1965}, which allows simulating heat transport within the GK formalism, using equilibrium AIMD. The second difficulty was addressed using advanced statistical methods for the spectral analysis of stationary time series, which provide an asymptotically unbiased and consistent estimate of the power spectrum of the energy current, meaning that both the bias and the statistical error of the estimated conductivity can be made arbitrarily and controllably small in the limit of long simulation time \citep{Ercole2017}. The combination of these theoretical and methodological innovations makes the \emph{ab initio} simulation of heat transport feasible in principle and affordable in practice. In this chapter we review the efforts that have led to these achievements, starting with a brief review of the relevant theoretical concepts, and illustrate them with relevant benchmarks and an application to a realistic model of liquid water.

%%%%%%%%%%%%%%%%%%%%%%%%%%%%%%%%%%%%%%%%%%%%%%%%%%%%%%%%%%%%%%%%%%%%%%%%%%%%%%%%%%%%
\section{Green-Kubo theory of heat transport}

\subsection{Hydrodynamic variables} \label{sec:hydrodyn_var}
The macroscopic processes occurring in condensed matter are often described in terms of \emph{extensive variables}. By definition, the value that such a variable assumes for a system is the sum of the values it has for each of its subsystems. This property allows one to express an extensive variable, $A$, as the integral of a suitably defined density, $a(\mathbf{r})$, as:
\begin{equation}
A[\rOmega]=\int_\rOmega a(\mathbf{r})d\mathbf{r}, \label{eq:extensivity}
\end{equation}
where $\rOmega$ is the system volume. Here and in the following boldfaces indicate 3D vectors and Greek subscripts label Cartesian components: $\mathbf{u}= \{u_\alpha\} = \{u_1,u_2,u_3\}$. When an extensive quantity is locally conserved, a current density, $\bm{j}(\mathbf{r},t)$, can be associated to its density in such a way that the two of them satisfy the continuity equation:
\begin{equation}
\frac{\partial a(\mathbf{r},t)}{\partial t} = - \nabla\cdot\bm{j}(\mathbf{r},t), \label{eq:continuity}
\end{equation}
where $\nabla\cdot\bm{j}$ indicates partial differentiation and the middle dot a scalar product (a divergence in this case). In the following the densities and current densities of conserved quantities will be called \emph{conserved densities} and \emph{conserved currents} for short. The space Fourier transform of Eq.~\eqref{eq:continuity} reads:
\begin{equation}
\dot{\tilde a}(\mathbf{q},t) = - i\mathbf{q} \cdot \tilde {\bm{\jmath}} (\mathbf{q},t), \label{eq:kontinuity}\end{equation}
where the overdot indicates a time derivative and the tilde a Fourier transform, so that the longer the wavelength, the slower is the dynamics of a conserved density. We conclude that for long enough wavelengths, conserved densities are adiabatically decoupled from all the other (zillions of) fast atomic degrees of freedom. Note that in this chapter we are using the concept of \emph{adiabatic decoupling} in two distinct senses, depending on the context: to indicate the decoupling of electronic from nuclear degrees of freedom, and that of hydrodynamic variables from fast atomic ones.

The long-wavelength Fourier components of conserved densities are called \emph{hydrodynamic variables}. In macroscopically homogeneous systems, different wavelengths are decoupled from each other, while, as we have seen, the long wavelengths are adiabatically decoupled from all the other degrees of freedom. Let us suppose there are $Q$ conserved extensive variables. In the case of a mono-atomic fluid, for instance, $Q=5$, corresponding to mass (or particle number), energy, and the three components of the momentum. In order to simplify the notation, we set the value of the conserved quantities equal to zero, $A^i=0$, so that their densities, $a^i(\mathbf{r})$, directly refer to the departure from equilibrium, and we indicate by $\bm j^i(\mathbf{r},t)$ the corresponding currents. At equilibrium, all the conserved densities and currents vanish. Off equilibrium, it will be assumed that the wavelength and the time scale of the disturbances are so long that thermal equilibrium still holds \emph{locally}. That is to say, a local temperature, pressure, and chemical potential can be defined, such that, when combined with the densities of extensive variable, they satisfy a local equation of state.

For small enough deviations from equilibrium, the time derivatives of conserved densities are linear combinations of the densities themselves. In the frequency/wavevector domains this condition can be expressed as
\begin{equation}
  -i\omega\tilde a^i(\mathbf{q},\omega) = \sum_j \tilde\rLambda^{ij}(\mathbf{q},\omega) \tilde a^j(\mathbf{q},\omega), \label{eq:Fourier-continuity}
\end{equation}
where the tilde indicates now a space-time Fourier transform: $\tilde a(\mathbf{q},\omega) = \int \mathrm{e}^{-i(\mathbf{q}\cdot \mathbf{r}-\omega t)} a(\mathbf{r},t)d\mathbf{r}dt $. By combining Eq.~\eqref{eq:Fourier-continuity} with the time Fourier transform of Eq.~\eqref{eq:kontinuity}, we obtain the so-called constitutive equations for the (longitudinal components of the) conserved currents:
\begin{equation}
  \tilde{\bm{\jmath}}^i(\mathbf{q},\omega)=i\frac{\mathbf{q}}{q^2} \sum_j\tilde \rLambda^{ij}(\mathbf{q},\omega)\tilde a^j(\mathbf{q},\omega). \label{eq:constitutive-qomega}
\end{equation}
In isotropic media, the $\tilde\rLambda$'s are spherically symmetric functions of $\mathbf{q}$, whereas their value at $\mathbf{q}=0$ vanishes, because a non-vanishing value would imply a non-physical long-range dependence of the currents on density fluctuations, in contrast with our assumption of local thermodynamic equilibrium. The long-wavelength low-frequency limit of the coupling constants can thus be assumed to be $\tilde\rLambda^{ij}(\mathbf{q},\omega) \sim q^2 \lambda^{ij}$, so that the macroscopic ($\mathbf{q}=0$) stationary ($\omega=0$) components of the currents, $\mathbf{J}^i = \frac{1}{\rOmega} \int\bm j^i (\mathbf{r}) d\mathbf{r}$, are related to the corresponding components of the density gradients, $\mathbf{D}^i=\frac{1}{\rOmega}\int\nabla a^i(\mathbf{r})d\mathbf{r}$, through the equations:
\begin{equation}
  \mathbf{J}^i=\sum_j \lambda^{ij}\mathbf{D}^j. \label{eq:constitutive}
\end{equation}
In the following, the macroscopic component of a current will be indicated as a \emph{flux}.

Let $x^i=\frac{\partial S}{\partial A^i}$ be the intensive variable conjugate to $A^i$, where $S$ is the system's entropy, and $\chi^{ij} = \frac{1}{\rOmega} \frac{\partial A^i}{\partial x^j}$ the corresponding susceptibility. For instance, when $A^i$ is the energy of the system, the corresponding conjugate variable is the inverse temperature, $x^i=1/T$, while, when $A^i$ represents the number of particles of a given species, one has $x^i= - \mu^i/T$, $\mu^i$ being the corresponding chemical potential. The hypothesis of local thermodynamic equilibrium allows defining local values of the intensive variables, and we define \emph{thermodynamic forces} as their average gradients: $\mathbf{F}^i= \frac{1}{\rOmega} \int \nabla x^i(\mathbf{r})d\mathbf{r}$. The average density gradients are related to the thermodynamic forces through the susceptibility defined above, as:
\begin{equation}
\mathbf{D}^i=\sum_j\chi^{ij}\mathbf{F}^j .
\end{equation}
By inserting this relation into Eq.~\eqref{eq:constitutive}, one gets:
\begin{equation}
\mathbf{J}^i=\sum_j L^{ij} \mathbf{F}^j, \label{eq:onsager}
\end{equation}
where $L^{ij}=\sum_k\lambda^{ik}\chi^{kj}$. Eq.~\eqref{eq:onsager} expresses the linear relation between fluxes, the $\mathbf{J}$'s, and thermodynamic affinities, the $\mathbf{F}$'s, for which Onsager derived his celebrated reciprocity relations ($L^{ji}=L^{ij}$) from microscopic reversibility \citep{Onsager1931a,Onsager1931b,Casimir1945}. Note that, according to our definition, both the $\mathbf{J}$'s and the $\mathbf{F}$'s in Eq.~\eqref{eq:onsager} do not depend on the size of the system.

\subsection{Linear-response theory}
In order to evaluate the $L^{ij}$ phenomenological coefficients appearing in Eq.~\eqref{eq:onsager}, we consider a classical system of $N$ interacting atoms described by the Hamiltonian
\begin{equation}
  H^\circ(\rGamma) = \sum_n\frac{1}{2M_n}(\mathbf{P}_n)^2 + V(\mathbf{R}_1,\mathbf{R}_2,\cdots \mathbf{R}_N), \label{eq:unperturbed_H}
\end{equation}
where $M_n$, $\mathbf{R}_n$, and $\mathbf{P}_n$ are the masses, coordinates, and momenta of the $n$-th particle, $\rGamma=\{\mathbf{R}_n,\mathbf{P}_n\}$ indicates the phase-space coordinates of the entire system, and $V$ is a generic many-body potential. Let us now suppose that the system is subject to an external perturbation that can be described as a linear combination of the conserved densities, $\{a^i(\mathbf{r};\rGamma)\}$, as:
\begin{equation}
   V'(\rGamma,t) = \sum_i \int  v^i(\mathbf{r},t) a^i(\mathbf{r};\rGamma) d\mathbf{r}, \label{eq:perturbation}
\end{equation}
where $a(\mathbf{r};\rGamma)$ is a phase-space function whose ensemble average is the conserved density,
\begin{equation}
    \begin{aligned}
      a(\mathbf{r}) &= \langle a(\mathbf{r};\rGamma) \rangle \\
      & = \int a(\mathbf{r};\rGamma) \mathcal{P}^\circ(\rGamma)d\rGamma,
    \end{aligned}
\end{equation}
$\mathcal{P}^\circ(\rGamma) \propto \mathrm{e}^{-\frac{H^\circ(\rGamma)}{k_BT}}$ is the equilibrium distribution, $k_B$ the Boltzmann constant, and $\{v^i(\mathbf{r},t)\}$ are time-dependent fields that couple to the conserved densities and vanish at $t=-\infty$, when the system is assumed to be in thermal equilibrium at some temperature $T$. Of course, conserved currents are also expected values of some phase-space functions, $\bm{j}(\mathbf{r})=\langle \bm{j}(\mathbf{r};\rGamma)\rangle$. The phase-space functions whose expected values are conserved densities/currents will be referred to as \emph{phase-space samples} of the currents/densities. In the following, when the phase-space dependence of a conserved density/current is explicitly indicated, we will mean a phase-space sample; when it is not a phase-space average will be implied. When a phase-space sample is evaluated along a dynamical trajectory, $\rGamma_t$, the sample function will depend on time and on the initial conditions of the trajectory. Averaging with respect to the initial conditions will result in a time-dependent expected value for the conserved densities (or currents):
\begin{equation}
  \begin{aligned}
    a(\mathbf{r},t) &= \langle a(\mathbf{r};\rGamma'_t)\rangle_0 \\
    &= \int a(\mathbf{r};\rGamma'_t) \mathcal{P}^\circ(\rGamma_0) d\rGamma_0.
  \end{aligned} \label{eq:a(r,t)}
\end{equation}
In Eq.~\eqref{eq:a(r,t)} the notation $\rGamma'_t$ denotes somewhat pedantically that the time evolution in phase space is driven by the perturbed Hamiltonian, $H^\circ+V'$. If it were driven by $H^\circ$, evidently the value of $a$ would be time-independent. In the following, the notation $\rGamma_t$ will indicate an unperturbed time evolution. As an example, the phase-space sample of the particle density can be assumed to be $n(\mathbf{r};\rGamma) = \sum_n \delta (\mathbf{r}-\mathbf{R}_n)$, the corresponding current is $\bm{j} (\mathbf{r}, \rGamma) = \sum_n \delta(\mathbf{r}-\mathbf{R}_n) \mathbf{P}_n / M_n $, and a local external potential is described by: $V'(\rGamma,t)=\sum_n v(\mathbf{R}_n,t)=\int v'(\mathbf{r},t) n(\mathbf{r};\rGamma) d \mathbf{r}$. Note that sample functions are not necessarily univocally defined. Different functions whose phase-space averages coincide in the long-wavelength limit sample the same hydrodynamical variable. More on this in Sec.~\ref{sec:gauge-invariance}.

According to \cite{Green1954}, \cite{Kubo1957a}, and \cite{Kubo1957b}, the linear response of the $i$-th conserved current to the perturbation is:
\begin{align}
  j_\alpha^i(\mathbf{r},t) & = \frac{1}{k_B T} \sum_j \int_{-\infty}^t dt' \int d\mathbf{r}' \Bigl \langle j_\alpha^i(\mathbf{r},\rGamma_t)\dot a^j(\mathbf{r}',\rGamma_{t'})\Bigr \rangle_0 v^j(\mathbf{r}',t') \\
  &= \frac{-1}{k_B T} \sum_{j,\beta} \int_{-\infty}^t dt' \int d\mathbf{r}' \Bigl \langle j_\alpha^i(\mathbf{r},\rGamma_t) \partial'_\beta j_\beta^j(\mathbf{r}',\rGamma_{t'})\Bigr \rangle_0 v^j(\mathbf{r}',t') \\
  &= \frac{1}{k_B T} \sum_{j,\beta} \int_{-\infty}^t dt' \int d\mathbf{r}' \left \langle j_\alpha^i(\mathbf{r},\rGamma_{t}) j_\beta^j(\mathbf{r}',\rGamma_{t'})\right \rangle_0 \partial'_\beta v^j(\mathbf{r}',t'). \label{eq:linear-response-c}
\end{align}
The second line follows from the first through the continuity equation, Eq.~\eqref{eq:continuity}, while the third line follows after integrating by parts with respect to $\mathbf{r}'$. The notation $\partial'_\beta=\frac{\partial}{\partial r'_\beta}$ has been used.

By integrating Eq.~\eqref{eq:linear-response-c} all over the space, and assuming space-time homogeneity as well as isotropy, one recovers Eq.~\eqref{eq:onsager} with:
\begin{align}
J^i_\alpha(\rGamma) &= \frac{1}{\rOmega} \int j^i_\alpha(\mathbf{r},\rGamma) d\mathbf{r}, \label{eq:J_def}\\
F^i_\alpha(\rGamma) &= \frac{1}{\rOmega T} \int \partial_\alpha v^i(\mathbf{r},\rGamma) d\mathbf{r}, \label{eq:F_def}\\
L^{ij}_{\alpha\beta} &= \frac{\rOmega}{k_B} \int_0^\infty \left\langle J^i_\alpha(\rGamma_t) J^j_\beta(\rGamma_0)\right\rangle_0 dt. \label{eq:L_def}
\end{align}
This completes the derivation of the Green-Kubo formula for transport coefficients, Eq.~\eqref{eq:GK}, from classical linear-response theory. Onsager's reciprocity relations, $L^{ij}=L^{ji}$ \citep{Onsager1931a,Onsager1931b}, follow from Eq.~\eqref{eq:L_def} leveraging time-translational invariance, $\langle J^i_\alpha(\rGamma_t) J^j_\beta(\rGamma_0) \rangle = \langle J^i_\alpha(\rGamma_0) J^j_\beta(\rGamma_{-t}) \rangle$, and micro-reversibility, $\langle J^i_\alpha(\rGamma_t) J^j_\beta(\rGamma_0) \rangle = \langle J^i_\alpha(\rGamma_{-t}) J^j_\beta(\rGamma_0) \rangle$.

\subsubsection{Einstein-Helfand expression for transport coefficients and the Wiener-Khintchine theorem}  \label{sec:Einstein}
The celebrated Einstein's relation between the mean-square displacement of a diffusing particle and its velocity auto-correlation function is easily generalized to an arbitrary stochastic process and has in fact been utilized by \cite{Helfand1960} to provide an ``Einstein-like'' expression for transport coefficients.

Let $X_t$ be a stationary stochastic process. One has:
\begin{equation}
  \frac{1}{\mathcal{T}} \left \langle \left | \int_0^\mathcal{T} X_t dt \right |^2 \right \rangle = 2 \int_0^\mathcal{T} \left \langle X_t X_0 \right \rangle dt -\frac{2}{\mathcal{T}} \int_0^\mathcal{T} \left \langle X_t X_0 \right \rangle t \,dt. \label{eq:Einstein-Helfand}
\end{equation}
In the large-$\mathcal{T}$ limit, the second term on the right-hand side of Eq.~\eqref{eq:Einstein-Helfand} can be neglected.

When the stochastic process is the velocity of a Brownian particle, Eq.~\eqref{eq:Einstein-Helfand} allows one to establish a relation between the diffusion constant of the particle, temperature, and the auto-correlation time of the velocity.  When $X_t$ is the heat flux of a macroscopic body, Eq.~\eqref{eq:Einstein-Helfand} allows one to estimate the thermal conductivity, as given by Eq.~\eqref{eq:GK}, from the asymptotic behavior of the ``energy displacement'' $\mathcal{D}(\tau) = \int_0^\tau \mathbf{J}(\rGamma_t) dt $.

Eq.~\eqref{eq:Einstein-Helfand} can be easily generalized to the finite-frequency regime, to get:
\begin{equation}
  \begin{aligned}
    S_\mathcal{T}(\omega) &= \frac{1}{\mathcal{T}} \left \langle \left | \int_0^\mathcal{T} X_t \mathrm{e}^{i\omega t}dt \right |^2 \right \rangle \\
    &= 2\mathfrak{Re} \int_0^\mathcal{T} \left \langle X_t X_0 \right \rangle \mathrm{e}^{i\omega t}dt + \mathcal{O}(\mathcal{T}^{-1}).
  \end{aligned}
  \label{eq:Wiener-Khintchine}
\end{equation}
This equation expresses the Wiener-Khintchine theorem \citep{Wiener1930,Khintchine1934}, which states that the expectation of the squared modulus of the Fourier transform of a stationary process is the Fourier transform of its time correlation function, which is usually referred to as the process \emph{power spectral density},
\begin{equation}
  S(\omega) = \int_{-\infty}^\infty \langle X_t X_0 \rangle \,\mathrm{e}^{i\omega t} dt, \label{eq:S(omega)}
\end{equation}
aka the \emph{power spectrum}. In the following the suffix $\mathcal{T}$ will be neglected for simplicity and its value assumed to be sufficiently large as to be considered infinite. More generally, when several conserved currents interact with each other, one can define the \emph{cross-spectrum} of the conserved fluxes as the Fourier transform of the cross time-correlation functions:
\begin{equation}
  \begin{aligned}
    S^{kl}(\omega) &= \int_{-\infty}^\infty \langle X^k_t X^l_0 \rangle \,\mathrm{e}^{i\omega t} dt \\
    &= \frac{1}{\mathcal{T}} \mathfrak{Re} \left\langle \int_0^\mathcal{T} X^k_t \mathrm{e}^{-i\omega t}dt \times \int_0^\mathcal{T} X^l_t \mathrm{e}^{i\omega t}dt \right\rangle + \mathcal{O}(\mathcal{T}^{-1}).
  \end{aligned} \label{eq:Sij(omega)}
\end{equation}
Eqs.~\eqref{eq:Einstein-Helfand} and \eqref{eq:Wiener-Khintchine} indicate that the transport coefficients we are after essentially are the zero-frequency value of the (cross-) power spectrum of the corresponding current(s), a fact that will be instrumental in our approach to data analysis, as explained in Sec.~\ref{sec:data-analysis}. Therefore, Eq.~\eqref{eq:L_def} can be cast into the form:
\begin{equation}
    L^{kl} =\frac{\rOmega}{2 k_B} S^{kl}(\omega=0), \label{eq:GK-S0}
\end{equation}
where the Cartesian indices have been omitted for clarity.

\subsection{Heat transport}
The above treatment allows one to compute the linear response of a system at thermal equilibrium to a generic mechanical perturbation. Heat transport is determined by temperature gradients that cannot be described by any mechanical perturbation. The concept of temperature distribution implies that the system is locally at thermal equilibrium over lengths and times large with respect to atomic distances and relaxation times. Temperature affects the physical properties of a system through the Boltzmann distribution function. When the temperature is not constant, $T(\mathbf{r})=T+\rDelta T(\mathbf{r})$ ($|\rDelta T| \ll T$), the effects of this inhomogeneity can be formally described by the distribution function:
\begin{align}
  \mathcal{P}(\rGamma) & \propto
  \mathrm{e}^{-\int \frac{e(\mathbf{r};\rGamma)}{k_BT(\mathbf{r})}d\mathbf{r}} \\
  &= \mathrm{e}^{-\frac{H^\circ(\rGamma)+V'(\rGamma)}{k_BT}},
\end{align}
where $e(\mathbf{r};\rGamma)$ is an energy (Hamiltonian) density, such that $\int e(\mathbf{r};\rGamma) d\mathbf{r}= H^\circ(\rGamma)$. Eq.~\eqref{eq:perturbation} becomes:
\begin{equation}
   V'(\rGamma)=-\frac{1}{T}\int\rDelta T(\mathbf{r})e(\mathbf{r};\rGamma) d\mathbf{r} + \mathcal{O}(\rDelta T^2) . \label{eq:mechanical-perturbation}
\end{equation}
Eq.~\eqref{eq:mechanical-perturbation} shows that the effects of temperature inhomogeneities can be mimicked by a mechanical perturbation coupled to the temperature distribution. From Eqs.~\eqref{eq:onsager} and (\ref{eq:J_def}-\ref{eq:L_def}) we conclude that in a system where the only non-trivial conserved quantity is the energy, the heat (energy) flow is coupled to temperature gradients through the constitutive equation:
\begin{equation}
  \mathbf{J}^\smallE = - \kappa \nabla T, \label{eq:thermal-constitutive}
\end{equation}
where the thermal conductivity $\kappa_{\alpha\beta}= L^{\scriptscriptstyle EE}_{\alpha \beta} / T^2$ (see Eq.~\eqref{eq:onsager}) can be expressed by a Green-Kubo relation in terms of the fluctuations of the energy flux as:
\begin{equation}
  \kappa_{\alpha\beta} =\frac{\rOmega}{k_B T^2} \int_0^\infty \left \langle J^\smallE_\alpha(\rGamma_t) J^\smallE_\beta(\rGamma_0) \right \rangle_0 dt, \label{eq:GK-complete}
\end{equation}
and
\begin{equation}
  \mathbf{J}^\smallE(\rGamma)  = \frac{1}{\rOmega} \int \bm{j}^\smallE(\mathbf{r};\rGamma) d\mathbf{r}. \label{eq:avg-current-density}
\end{equation}
In order to obtain an explicit expression for the energy flux from a microscopic expression for the energy density, we multiply the continuity equation, Eq.~\eqref{eq:continuity}, by $\mathbf{r}$ and integrate by parts, to obtain:
\begin{align}
  \mathbf{J}^\smallE(\rGamma_t) &= \frac{1}{\rOmega} \int  \dot e(\mathbf{r};\rGamma_t) \,\mathbf{r}\, d\mathbf{r} \label{eq:JE=rdote} \\
  & = \frac{1}{\rOmega} \int
  \left[ \sum_n \left(
    \frac{\partial e(\mathbf{r};\rGamma_t)}{\partial \mathbf{R}_n} \cdot \mathbf{V}_n +
    \frac{\partial e(\mathbf{r};\rGamma_t)}{\partial \mathbf{P}_n} \cdot \mathbf{F}_n
  \right)  \right] \mathbf{r}\, d\mathbf{r}, \label{eq:JE}
\end{align}
where $ \mathbf{F}_n $ is the force acting on the $n$-th atom, and $\mathbf{V}_n=\frac{\mathbf{P}_n}{M_n}$ its velocity.

The manipulations leading from the continuity equation, Eq.~\eqref{eq:continuity}, to Eq.~\eqref{eq:JE} deserve some further comments, as they imply neglecting a boundary term, $\mathbf{J}_{\partial \rOmega} =
\frac{1}{\rOmega} \int_{\partial\rOmega}
 \left( \bm{j}(\mathbf{r}) \cdot \hat{\mathbf{n}} \right)
\mathbf{r} \,d\mathbf{r} $  (where $ \partial\rOmega $ is the boundary of the integration volume and $\hat{\mathbf{n}}$ the normal to it), which in general does not vanish in the thermodynamic limit and is ill-defined in periodic boundary conditions (PBC). The correct way of addressing this problem is to work with the Taylor expansion of the space Fourier transform of the continuity equation, Eq.~\eqref{eq:kontinuity}, and to perform the thermodynamic limit at finite wavelength. The leading non-vaninishing term in the Taylor expansion yields Eq.~\eqref{eq:JE=rdote} without any boundary term in the way.

\subsubsection{Energy flux from classical force fields}
When atoms interact through a classical force field, $V(\mathbf{R}_1,\mathbf{R}_2,\cdots \mathbf{R}_N)$, an energy density can be defined in terms of local atomic energies as:
\begin{align}
  e(\mathbf{r},\rGamma) &= \sum_n \delta(\mathbf{r}-\mathbf{R}_n) e_n(\rGamma), \label{eq:epsilon-classical} \\
  e_n(\rGamma) &= \frac{(\mathbf{P}_n)^2}{2M_n} + v_n(\{\mathbf{R}\}), \label{eq:atomic-energies}
\end{align}
where the $v_n$'s are a set of atomic potential energies whose sum is the total potential energy of the system, $\sum_n v_n=V$, with a short-range dependence on the coordinates of the other atoms. In the presence of long-range forces, this condition is effectively guaranteed by local charge neutrality, which we will assume throughout.
By inserting Eq.~\eqref{eq:epsilon-classical} into Eq.~\eqref{eq:JE}, the energy flux can be cast into the form:
\begin{align}
  \mathbf{J}^\smallE(\rGamma) &=
  \frac{1}{\rOmega} \left[ \sum_n \mathbf{V}_n e_n + \sum_n \mathbf{R}_n
        \left( \mathbf{F}_n \cdot \mathbf{V}_n  + \sum_m \mathbf{V}_m \cdot {\frac{\partial v_n}{\partial \mathbf{R}_m}} \right) \right] \nonumber \\
       &= \frac{1}{\rOmega} \left[ \sum_n \mathbf{V}_n e_n + \sum_{n,m} (\mathbf{R}_n-\mathbf{R}_m) \mathbf{F}_{n m} \cdot \mathbf{V}_n \right], \label{eq:J-classical}
\end{align}
where $\mathbf{F}_{n m} = - \frac{\partial v_m}{\partial \mathbf{R}_n}$ is the contribution of the $m$-th atom to the force acting on the $n$-th atom, $\sum_m \mathbf{F}_{n m} = \mathbf{F}_{n}$, and $\mathbf{F}_{n m} = -\mathbf{F}_{m n}$. When the interaction amongst atoms can be expressed in terms of two-body potentials, one has: $v_m=\frac{1}{2}\sum_n v(\mathbf{R}_n- \mathbf{R}_m)$ and $\mathbf{F}_{n m} = - \frac{1}{2} \nabla_{\mathbf{R}_n} v(\mathbf{R}_n- \mathbf{R}_m)$. Here we implicitly assumed that the interaction energy is equally partitioned between atoms $m$ and $n$. In Sec.~\ref{sec:gauge-invariance} we shall see this is not the only possible choice, with far-reaching consequences on the theory of heat transport.

The first term on the right-hand side of Eq.~\eqref{eq:J-classical} is often called \emph{convective} and the second \emph{virial}. We feel that the wording ``convective'' is somewhat misleading in this context, as the convective current, as well as its contribution to heat conductivity, may not vanish even in the absence of convection.

\subsubsection{Multi-component fluids} \label{sec:multi-component}
In a multi-component fluid there is one conserved quantity (the particle number) per atomic species, plus the total energy and the three Cartesian components of the total momentum. The momentum densities are mass currents: the mass flux is therefore the total momentum, which vanishes in the center of mass reference frame. The transverse components of the momentum densities are decoupled from the other conserved densities \citep{Foster1975}, while the longitudinal one can be assumed to coincide with the total momentum in the long-wavelength limit. Momentum conservation thus constrains the number of fluxes interacting with the energy flux in Eq.~\eqref{eq:onsager} to $Q-1$, $Q$ being the number of atomic species, so that the resulting dimension of the matrix of Onsager coefficients, $L$, is $Q\times Q$. The heat flux is defined as the non-convective component of the energy flux, \emph{i.e.} the value of the latter in the absence of mass transport, that is to say when all the particle fluxes vanish.\footnote{It is unfortunate, but inevitable due to common usage, that this definition of non-convective flux clashes with a different definition given above while commenting Eq.~\eqref{eq:J-classical}.} By imposing this condition in Eq.~\eqref{eq:onsager}, with $\mathbf J^\smallone \equiv \mathbf{J}^\smallE$, and $\mathbf{J}^{q}$ ($q =2,\dots Q$) being independent particle fluxes, the thermal conductivity, defined as the ratio of the heat flux over the temperature gradient, is given by:
\begin{equation}
\kappa = \frac{1}{T^2 \,(L^{-1})^{\smallone\smallone}}. \label{eq:multi_kappa}
\end{equation}
This expression can be proved to be invariant under \textit{any} non-singular linear transformation of the independent particle fluxes. For instance, in the case of a two-component liquid, energy and particle currents are coupled as in:
\begin{equation}
  \begin{aligned}
    \mathbf{J}^\smallE &= L^{\smallE\smallE}\,  \nabla \left (\frac{1}{T} \right ) + L^{\smallE\smallQ} \, \nabla \left (\frac{\mu}{T} \right ), \\
    \mathbf{J}^{\smallQ} &= L^{\smallE\smallQ} \, \nabla \left (\frac{1}{T} \right ) + L^{\smallQ\smallQ} \, \nabla \left (\frac{\mu}{T} \right ), \label{eq:two-comp-constitutive}
  \end{aligned}
\end{equation}
where $\mathbf{J}^{\smallQ}$ is the particle current of one of the two species (say, the second), and $\mu$ the corresponding chemical potential \citep{Sindzingre1990}. By imposing that the particle current vanishes, the resulting thermal conductivity is:
\begin{equation}
  \kappa=\frac{1}{T^2}
  \left( L^{\smallE\smallE} - \frac{(L^{\smallE\smallQ})^2}{L^{\smallQ\smallQ}} \right). \label{eq:two-comp-kappa}
\end{equation}

%%%%%%%%%%%%%%%%%%%%%%%%%%%%%%%%%%%%%%%%%%%%%%%%%%%%%%%%%%%%%%%%%%%%%%%%%%%%%%%%%%%%
\section{Gauge invariance of heat transport coefficients} \label{sec:gauge-invariance} 
\begin{figure}
\begin{minipage}{0.5\textwidth}
\centering \includegraphics{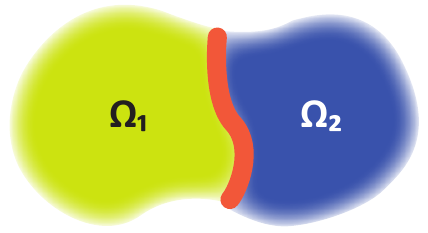}
\end{minipage}
\begin{minipage}{0.4\textwidth}
\begin{align*}
  E(\rOmega_{1}\cup\rOmega_2) &= E(\rOmega_1) + E(\rOmega_2) + W_{12} \qquad \\
  & \overset{?}{=}\mathcal{E}(\rOmega_1)+\mathcal{E}(\rOmega_2)
\end{align*}
\end{minipage}
\caption{
	The energy of an isolated system is the sum of the energies of its subsystems (as defined when they are isolated as well) plus the interaction among them, $W_{12}$, whose magnitude scales as the area of the interface, depicted in red. When defining the energies of individual subsystems, $\mathcal{E}$, $W_{12}$ has to be arbitrarily partitioned among them. \label{fig:energy-partition}}
\end{figure}

It is often implicitly assumed that the well-definiteness of thermal transport coefficients would stem from the uniqueness of the decomposition of the system's total energy into localized, atomic, contributions. This assumption is manifestly incorrect, as any decomposition leading to the same value for the total energy as Eq.~\eqref{eq:atomic-energies} should be considered as legitimate. The difficulty of partitioning a system's energy into subsystems' contributions is illustrated in Fig.~\ref{fig:energy-partition}, which depicts a system made of two interacting subsystems. When defining the energy of each of the two subsystems, an arbitrary decision has to be made as to how the interaction energy is partitioned. In the case depicted in Fig.~\ref{fig:energy-partition}, for instance, the energy of each of the two subsystems can be defined as $\mathcal{E}(\rOmega_i) = E(\rOmega_i) + \frac{1}{2}(1\pm\lambda)W_{12}$, where $E(\rOmega_i)$ are the energies of the two isolated subsystems, $W_{12}$ their interaction energy, and $\lambda$ an arbitrary constant. In the thermodynamic limit, when all the subsystems' energies are much larger than the interaction between any pairs of them, the value of the $\lambda$ constant is irrelevant. When it comes to defining energy densities (\emph{i.e.} energies of infinitesimal portions of a system) or atomic energies, instead, the magnitude of the interaction between different subsystems is comparable to their energies, which become therefore intrinsically ill-defined.

Let us consider a mono-atomic fluid interacting through pair potentials, $v(|\mathbf{R}_n-\mathbf{R}_m|)$, and define the atomic energies as \citep{Marcolongo2014,Ercole2016}:
\begin{equation}
  e_{\smallgamma,n}(\rGamma) =
  \frac{1}{2M_n}(\mathbf{P}_{n})^{2} + \frac{1}{2}\sum_{m\ne n}
  v(|\mathbf{R}_{n}-\mathbf{R}_{m}|)
  (1+\gamma_{nm}), \label{eq:Lambda-atomic-energies}
\end{equation}
where $\gamma_{nm}=-\gamma_{mn}$ is \emph{any} antisymmetric matrix.
As the inter-atomic potential appearing in Eq.~\eqref{eq:Lambda-atomic-energies} is symmetric with respect to the atomic indices, it is clear that the sum of all the atomic energies does not depend on $\gamma$, thus making any choice of $\gamma$ equally permissible. This trivial observation has deep consequences on the theory of thermal fluctuations and transport, because the value of the macroscopic energy flux, instead, depends explicitly on $\gamma$, thus making one fear that the resulting transport coefficients would depend on $\gamma$ as well. Using the same manipulations that lead from Eqs.~\eqref{eq:epsilon-classical} and \eqref{eq:atomic-energies} to Eq.~\eqref{eq:J-classical}, for any choice of the $\gamma$ matrix in Eq.~\eqref{eq:Lambda-atomic-energies}, a corresponding expression for the macroscopic energy flux can be found, reading \citep{Marcolongo2014,Ercole2016}:
\begin{equation}
  \mathbf{J}_{\smallgamma}^\smallE=\mathbf{J}^\smallE+
  \frac{1}{2 \rOmega}\sum_{n,m\ne n}\gamma_{nm} \Bigl ( v_{nm} \mathbf{V}_{n} + \bigl (\mathbf{V}_{n}\cdot \nabla_{\mathbf{R}_n} v_{nm} \bigr )  (\mathbf{R}_{n}-\mathbf{R}_{m}) \Bigr ), \label{eq:Lambda-classical-current}
\end{equation}
where $v_{nm}= v(|\mathbf{R}_{n} - \mathbf{R}_{m}|)$.

\begin{figure}
    \centering
    \begin{subfigure}[tb]{1\textwidth}
    	\centering
        \includegraphics[width=8cm]{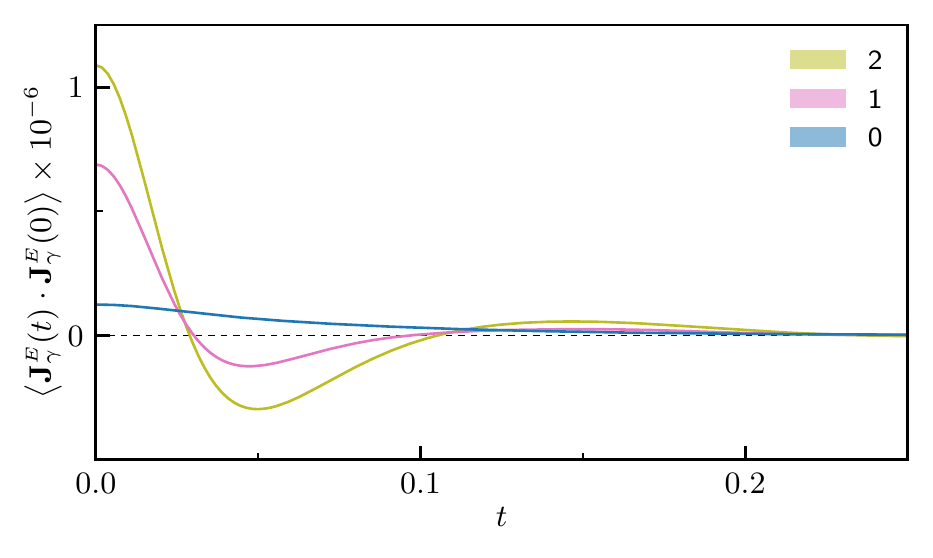}
        \caption{}
        \label{fig:argon-gauge-acf}
    \end{subfigure}

    \begin{subfigure}[tb]{1\textwidth}
    	\centering
        \includegraphics[width=8cm]{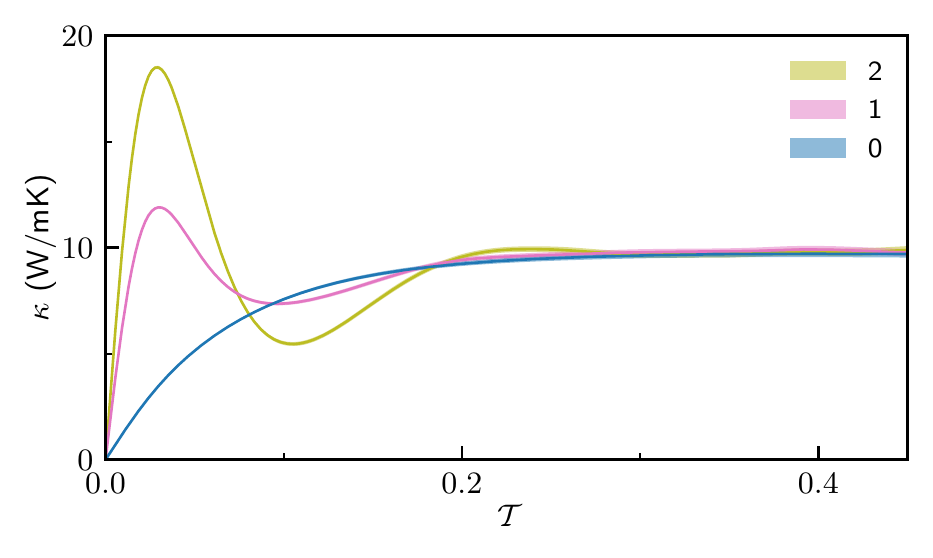}
        \caption{}
        \label{fig:argon-gauge-kappa}
    \end{subfigure}
	\caption{(a) Time correlation functions of the modified macroscopic energy flux of a Lennard-Jones fluid, at the conditions described in the text, as defined in Eq.~\eqref{eq:Lambda-classical-current}, for different definitions of the $\gamma$ matrix. The ``0'' line refers to the standard definition ($\gamma = 0$), whereas the labels ``1'' and ``2'' correspond to two other (arbitrary) definitions of $\gamma$ as described in \cite{Ercole2016}.
    (b) Integral of the time correlation functions displayed in Fig.~\ref{fig:argon-gauge-acf}, multiplied by the prefactor appearing in the GK relation, Eq.~\eqref{eq:GK-complete}, as a function of the upper limit of integration. The barely visible shaded area surrounding each line is an indication of the error bars, as estimated by standard block analysis. Units are Lennard-Jones units ($M=\sigma=\varepsilon=1$). \label{fig:argon-gauge}}
\end{figure}

As a specific example, \cite{Ercole2016} ran MD simulations for a Lennard-Jones monoatomic fluid described by the inter-atomic potential $v(r)=\epsilon \left [ \left ( \frac{\sigma}{r}\right )^{12} - \left ( \frac{\sigma}{r}\right )^{6}\right ] $ at temperature $T=1.86 \frac{\epsilon}{k_B}$ and density $\rho=0.925 \sigma^{-3}$. In Fig.~\ref{fig:argon-gauge-acf} we display the resulting macroscopic energy-flux autocorrelation function corresponding to different choices of the $\gamma$ matrix in Eqs.~\eqref{eq:Lambda-atomic-energies} and \eqref{eq:Lambda-classical-current}. Fig.~\ref{fig:argon-gauge-acf} clearly shows that the $\langle \mathbf{J}^\smallE_{\gamma}(t) \cdot\mathbf{J}^\smallE_{\gamma}(0) \rangle$ correlation functions dramatically depend on the $\gamma$ matrices in Eqs.~\eqref{eq:Lambda-atomic-energies} and \eqref{eq:Lambda-classical-current}.  Notwithstanding, the integrals of all these time correlation functions tend to the same limit at large integration times, as shown in Fig.~\ref{fig:argon-gauge-kappa}.

In order to get insight into this remarkable invariance property, let us inspect the difference between the generalized flux in Eq.~\eqref{eq:Lambda-classical-current} and the standard expression of Eq.~\eqref{eq:J-classical}:
\begin{equation}
  \rDelta\mathbf{J}^\smallE_{\smallgamma} =\mathbf{J}^\smallE_{\smallgamma}-\mathbf{J}^\smallE  =\frac{\mathrm{d}}{\mathrm{dt}}\frac{1}{4 \rOmega} \sum_{n,m\ne n}  \gamma_{nm} \, v(|\mathbf{R}_{n}-\mathbf{R}_{m}|)  (\mathbf{R}_{n}-\mathbf{R}_{m}). \label{eq:DeltaJ}
\end{equation}
We see that the two different expressions for the macroscopic energy flux differ by a total time derivative of a bounded phase-space vector function. In the following, we show that this is a consequence of energy conservation and extensivity and a sufficient condition for the corresponding thermal conductivities to coincide.

The very possibility of defining an energy current density, from which the energy fluxes of Eq.~\eqref{eq:J-classical} and \eqref{eq:Lambda-classical-current} ultimately depend, stems from energy extensivity. The considerations illustrated in Fig.~\ref{fig:energy-partition} indicate that any two densities, $e'(\mathbf{r},t)$ and $e(\mathbf{r},t)$, whose integrals over a macroscopic volume differ by a quantity that scales as the volume boundary, should be considered as equivalent. This equivalence can be expressed by the condition that two equivalent densities differ by the divergence of a (bounded) vector field:
\begin{equation}
  e'(\mathbf{r},t)=e(\mathbf{r},t) - \nabla\cdot \bm{p}(\mathbf{r},t). \label{eq:gauge_transformation}
\end{equation}
In a sense, two equivalent energy densities can be thought of as different \emph{gauges} of the same scalar field. Energy is also conserved: because of this, for any given gauge of the energy density, $e(\mathbf{r},t)$, an energy current density can be defined, $\bm{j}(\mathbf{r},t)$, so as to satisfy the continuity equation, Eq.~\eqref{eq:continuity}. By combining Eqs.~\eqref{eq:gauge_transformation} and \eqref{eq:continuity} we see that energy current densities and macroscopic fluxes transform under a gauge transformation as:
\begin{align}
 \bm{j}'(\mathbf{r},t) & = \bm{j}(\mathbf{r},t) + \dot{\bm{p}}(\mathbf{r},t), \label{eq:current_density_gauge} \\
  \mathbf{J}'(t) & = \mathbf{J}(t) + \dot{\mathbf{P}}(t), \label{eq:macroscopic_flux_gauge}
\end{align}
where $\mathbf{P}(t)=\frac{1}{\rOmega} \int\bm{p}(\mathbf{r},t)d\mathbf{r}$. We
conclude that the macroscopic energy fluxes in two different energy gauges differ by the total time derivative of a bounded phase-space vector function.

We now show that the energy fluxes of the same system in two different energy gauges, $e$ and $e'$, differing by a bounded total time derivative, as in Eq.~\eqref{eq:macroscopic_flux_gauge}, result in the same heat conductivity, as given by the Green-Kubo formula, Eq.~\eqref{eq:GK-complete}. More generally, the Onsager coefficients coupling two fluxes, $\mathbf{J}^1$ and $\mathbf{J}^2$, do not depend on the gauge of either one of them. In fact, let $\left(\mathbf{J}^1\right)' = \mathbf{J}^1 + \dot{\mathbf{P}}$; one has:
\begin{equation}
  \begin{aligned}
    \left (L^{11} \right)' &= \frac{\rOmega}{2 k_B }\int_{-\infty}^{+\infty} \left \langle \left (\mathbf{J}_1(t)+\dot{\mathbf{P}}(t) \right ) \cdot  \left (\mathbf{J}_1(0)+\dot{\mathbf{P}}(0)\right ) \right \rangle dt \\
    &= L^{11} + \frac{\rOmega}{2 k_B } \left[ \left .  \left \langle \mathbf{P}(t) \cdot \dot{\mathbf{P}}(0) \right \rangle \right |^{+\infty}_{-\infty} + \left .  2 \Bigl \langle \mathbf{P}(t) \cdot \mathbf{J}_1(0) \Bigr \rangle \right |^{+\infty}_{-\infty} \right] .
  \end{aligned} \label{eq:L=L'}
\end{equation}

The expectation of the time-lagged products in Eq.~\eqref{eq:L=L'} is equal to the products of two expectations at large time lag. As the equilibrium expectations of both a total time derivative and a current vanish, we conclude that $\left (L^{11}\right )'=L^{11}$. A slight generalization of this argument, also using microscopic reversibility as in \cite{Onsager1931a,Onsager1931b}, allows us to conclude that $\left (L^{12} \right )'=L^{12}$ and that, in general, $\kappa'=\kappa$.

\subsection{Molecular fluids} \label{sec:MolecularFluids}
In a one-component molecular fluid such as liquid water or, say, ethanol, there are in general $Q$ fluxes interacting with each other through Onsagers' Eq.~\eqref{eq:onsager}, where $Q$ is the number of atomic species in a molecule. The requirement that atoms are bound in molecules of fixed composition, however, sets a number of constraints that substantially simplify the treatment of heat transport, making the molecular case similar to the one-component one.

Let us consider a molecule of chemical formula $A_{N_A} B_{N_B}\cdots$, where $A, B,\cdots$ indicate atomic species, and $N_A,N_B,\cdots$ the corresponding atomic stoichiometric indices. For each atomic species we define the normalized number flux as:
\begin{equation}
  \mathbf{J}^X = \frac{1}{N_X}\sum_{n\in X} \mathbf{V}_n. \label{eq:JX}
\end{equation}
If we indicate by $M_X$ the atomic mass of species $X$, momentum conservation requires that $\sum_X M_X N_X \mathbf{J}^X = 0$ in the center-of-mass reference frame. The flux $\mathbf{J}^{XY} = \mathbf{J}^{X}-\mathbf{J}^{Y}$ is the total time derivative of a bounded vector, because its integral is the sum over all the molecules of the difference between the average atomic positions of either species within a same molecule, which is obviously bounded if molecules do not dissociate. As any number flux $\mathbf{J}^X$ can be expressed as a linear combination of the total momentum and of several $\mathbf{J}^{XY}$ fluxes, each of them is the total time derivative of a bounded vector. Therefore, the Onsager coefficient coupling any of these atomic fluxes with any other, or with the energy flux, vanishes. We conclude that energy is the only conserved quantity relevant for heat transport in a molecular fluid, and that the energy-flux autocorrelation function directly yields the thermal conductivity, as in Eq.~\eqref{eq:GK}.

%%%%%%%%%%%%%%%%%%%%%%%%%%%%%%%%%%%%%%%%%%%%%%%%%%%%%%%%%%%%%%%%%%%%%%%%%%%%%%%%%%%%
\section{Density-functional theory of adiabatic heat transport} \label{sec:DFT}
Quantum simulation methods based on Density-Functional Theory (DFT) have long been thought to be incompatible with the GK theory of thermal transport \emph{because in first-principles calculations it is impossible to uniquely decompose the total energy into individual contributions from each atom} \citep{Stackhouse2010b}. For this reason, \emph{ab initio} simulations of heat transport have often been performed using non-equilibrium approaches.

\cite{Stackhouse2010b}, for instance, computed the thermal conductivity of periclase MgO using a method devised by \cite{Müller-Plathe1997}. In this apporach a net heat flux, rather than a temperature gradient, is imposed to the simulated system and the thermal conductivity is evaluated as the ratio between the heat flux and the resulting temperature gradient.

In the so-called \emph{approach to equilibrium} methodology of \cite{Lampin2013} the system is first prepared in an out-of-equilibrium state characterized by an inhomogeneous temperature distribution and the thermal conductivity is evaluated from the time it takes for the system to relax to equilibrium. This technique has been combined with AIMD to simulate thermal transport in a GeTe$_4$ glass by \cite{Bouzid2017} and further generalized and applied to crystalline and nano-structured MgO by \cite{Puligheddu2017}.

Recently, there have been several attempts to combine the GK approach to heat transport with \emph{ab initio} techniques based on electronic-structure theory, by adopting some \emph{ad hoc} definitions for the energy flux. \cite{Kang2017}, for instance, derived an expression for the energy flux from a (rather arbitrary) quantum-mechanical definition of the atomic energies and used a modified MD integration algorithm to cope with the difficulties ensuing from the implementation of their expression in PBC. \cite{Carbogno:2017gc} gave a different expression for the energy flux, based on a normal-mode decomposition of the atomic coordinates and forces, which, while allowing to reduce the effects of thermal fluctuations, can only be applied to crystalline solids.

In spite of the undoubted ingenuity of these proposals, the problem still remains as of how it is possible that a rather arbitrary definition of the heat flux results in an allegedly well defined value for the thermal conductivity. The gauge-invariance principle introduced in Sec.~\ref{sec:gauge-invariance} not only provides a solution to this conundrum, but it also gives a rigorous way of deriving an expression for the energy flux directly from DFT, without introducing any \emph{ad hoc} ingredients.

In order to derive such an expression for the adiabatic energy flux, we start with the standard DFT expression of the total energy in terms of the Kohn-Sham (KS) eigenvalues $\varepsilon_v$, eigenfunctions $\phi_v(\mathbf{r})$, and density $n(\mathbf{r}) = \sum_v |\phi_v(\mathbf{r})|^2$ \citep{Martin2008}:
\begin{multline}
  E_{\smallDFT} = \frac{1}{2}\sum_{n}M_{n}V_{n}^{2} + \frac{\mathtt{e}^2}{2}\sum_{n,m\ne n}\frac{ Z_{n}Z_{m}}{|\mathbf{R}_{n}-\mathbf{R}_{m}|} \\
  + \sum_{v}\varepsilon_{v}-\frac{\mathtt{e}^2}{2}\int\frac{n(\mathbf{r})n(\mathbf{r}')}{|\mathbf{r}-\mathbf{r}'|}d\mathbf{r}d\mathbf{r}'+\int\left(\epsilon_{\smallXC}[n](\mathbf{r})-\mu_{\smallXC}[n](\mathbf{r})\right)n(\mathbf{r})d\mathbf{r},
\end{multline}
where $\mathtt{e}$ is the electron charge, $\epsilon_\smallXC[n](\mathbf{r})$ is a local exchange-correlation (XC) energy per particle defined by the relation $ \int \epsilon_\smallXC[n](\mathbf{r})n(\mathbf{r}) d\mathbf{r}=E_\smallXC [n]$, the latter being the total XC energy of the system, and $ \mu_\smallXC (\mathbf{r}) = \frac{\delta E_\smallXC }{\delta n(\mathbf{r})}$ is the XC potential. The DFT total energy can be readily written as the integral of a DFT energy density \citep{Chetty1992}:
\begin{equation}
  \begin{aligned}
    E_{\smallDFT} & =  \int e_{\smallDFT}(\mathbf{r})d\mathbf{r},\\
    e_{\smallDFT}(\mathbf{r}) & = e_{el}(\mathbf{r})+e_{\smallZ}(\mathbf{r}),
  \end{aligned}
  \label{eq:DFT-Edensity}
\end{equation}
where:
\begin{align}
  e_{el}(\mathbf{r}) & =\mathfrak{Re} \sum_{v}\phi_{v}^{*}(\mathbf{r})\bigl(H_{\smallKS}\phi_{n}(\mathbf{r})\bigr) \nonumber \\
  & \qquad\qquad\qquad - \frac{1}{2}n(\mathbf{r})v_{\smallH}(\mathbf{r}) +\left(\epsilon_\smallXC (\mathbf{r}) - \mu_\smallXC  (\mathbf{r}) \right) n(\mathbf{r}), \\
  e_{\smallZ}(\mathbf{r}) & = \sum_{n}\delta(\mathbf{r}-\mathbf{R}_{n}) \left(\frac{1}{2}M_{n}V_{n}^{2}+w_{n}\right), \\
  w_{n} & =\frac{\mathtt{e}^2}{2}\sum_{m\ne n}\frac{ Z_{n}Z_{m}}{|\mathbf{R}_{n}-\mathbf{R}_{m}|}, \label{eq:DFT-Edensity-breakup}
\end{align}
$H_\smallKS$ is the instantaneous self-consistent Kohn-Sham Hamiltonian, and $v_\smallH = \mathtt{e}^2 \int d\mathbf{r}' \frac{n(\mathbf{r}')}{|\mathbf{r}-\mathbf{r}'|}$ is the Hartree potential. An explicit expression for the DFT energy flux is obtained by computing the first moment of the time derivative of the energy density, Eqs.~(\ref{eq:DFT-Edensity}-\ref{eq:DFT-Edensity-breakup}), as indicated in Eq.~\eqref{eq:JE=rdote}, resulting in a number of terms, some of which are either infinite or ill-defined in PBC. Casting the result in a regular, boundary-insensitive, expression requires a careful breakup and refactoring of the various harmful terms, as explained by \cite{Marcolongo2014} and in the online version of \cite{Marcolongo2016}. The final result reads:
\begin{align}
  \allowdisplaybreaks
  \mathbf{J}^\smallE_{\smallDFT} &=\mathbf{J}^{\smallH} + \mathbf{J}^{\smallZ} + \mathbf{J}^{0} + \mathbf{J}^{\smallKS} +  \mathbf{J}^{\smallXC}, \\
  \mathbf{J}^{\smallH} &=
  \frac{1}{4\pi \rOmega \mathtt{e}^2}\int \nabla v_{\smallH}(\mathbf r) \dot v_{\smallH}(\mathbf  r) d\mathbf{r}, \\
  \mathbf{J}^{\smallZ} \label{eq:DFT-ionic} &=
  \frac{1}{\rOmega} \sum_{n}  \left[\mathbf{V}_{n}\left(\frac{1}{2}M_{n}V_{n}^{2} + w_{n}\right) + \sum_{m\ne n}(\mathbf{R}_{n} - \mathbf{R}_{m}) \left(\mathbf{V}_{m} \cdot \frac {\partial w_{n}}{\partial \mathbf{R}_{m}} \right) \right], \\
  \mathbf{J}^{0}  &=
  \frac{1}{\rOmega}  \sum_{n} \sum_{v}\left\langle \phi_{v} \left|(\mathbf{r}-\mathbf{R}_{n})\left(\mathbf{V}_{n}\cdot\frac{\partial\hat{v}_{0}}{\partial\mathbf{R}_{n}}\right)\right| \phi_{v}\right\rangle, \\
  \mathbf{J}^{\smallKS} &= \label{eq:DFT-KS}
  \frac{1}{\rOmega}  \mathfrak{Re} \sum_{v} \langle \bm{\bar \phi}_v^c | H_{\smallKS}+\varepsilon_v | \dot \phi_v^c \rangle, \\
  J_\alpha^{\smallXC} &=
  \begin{cases}
    0 & \mathrm{(LDA)} \\
      -\frac{1}{\rOmega} \int n(\mathbf{r}) \dot{n}(\mathbf{r}) \frac{\partial\epsilon^{\smallGGA} (\mathbf{r})}{\partial(\partial_\alpha n)} d\mathbf{r} & \mathrm{(GGA)},
  \end{cases}
\end{align}
where  $\hat v_0$ is the bare, possibly non-local, (pseudo-) potential acting on the electrons and
\begin{align}
  |\bm{\bar \phi}_v^c\rangle &= \hat P_c \,\mathbf{r}  \,|\phi_v \rangle, \label{eq:r-times-phi} \\
  |\dot \phi_v^c\rangle &= \dot{\hat P}_v \,|\phi_v\rangle, \label{eq:phi-dot}
\end{align}
are the projections over the empty-state manifold of the action of the position operator over the $v$-th occupied orbital, Eq.~\eqref{eq:r-times-phi}, and of its adiabatic time derivative \citep{Giannozzi2017}, Eq.~ \eqref{eq:phi-dot}, $\hat P_v$ and $\hat P_c = 1 - \hat P_v$ being the projector operators over the occupied- and empty-states manifolds, respectively. Both these functions are well defined in PBC and can be computed, explicitly or implicitly, using standard density-functional perturbation theory \citep{Baroni2001}.

%%%%%%%%%%%%%%%%%%%%%%%%%%%%%%%%%%%%%%%%%%%%%%%%%%%%%%%%%%%%%%%%%%%%%%%%%%%%%%%%%%%%
\section{Data analysis} \label{sec:data-analysis}
The MD evaluation of the GK integral, Eq.~\eqref{eq:GK}, usually proceeds in two steps. One first evaluates the integrand as a running average of the time-lagged current products, $\langle J^i(\tau)J^j(0)\rangle \sim \frac{1}{\mathcal{T}-\tau} \int_0^{\mathcal{T}-\tau} J^i(t+\tau)J^j(t)dt$, where $\mathcal{T}$ is the length of the MD trajectory.
The matrix defined in Eq.~\eqref{eq:L_def} is then estimated as a function of the upper limit of integration:  $L^{ij}(\mathcal{T})\propto \frac{\rOmega}{k_B} \int_0^\mathcal{T} \langle J^i(\tau)J^j(0)\rangle \,d\tau$. One then recovers, via Eq.~\eqref{eq:multi_kappa}, an estimate for the thermal conductivity depending on $\mathcal{T}$:
$\kappa(\mathcal{T})\propto \frac{1}{(L^{-1}(\mathcal{T}))^{\smallone\smallone}}$.
This function is usually very noisy: in fact, at times greater than the correlation time between $J^i$ and $J^j$, the correlation function $\langle J^i(\tau)J^j(0)\rangle$ approaches zero, hence $L^{ij}(\mathcal{T})$ starts integrating noise and behaves like the distance traveled by a random walk, whose variance grows linearly with the upper integration limit.
The evaluation of transport coefficients thus requires averaging over multiple trajectories (possibly multiple segments of a same long trajectory) and estimating the resulting uncertainty as a function of both the length of each trajectory and the upper limit of integration. This is a cumbersome task that often leads to a poor estimate of the statistical and systematic errors on the computed conductivity. All the more so when the signal is inherently oscillatory, due to the existence of high-frequency features in the power spectrum of the energy flux, possibly due to intramolecular oscillations that meddle with the noise. Some authors try to overcome these problems by either fitting the autocorrelation function or the GK integral with a multi-exponential function \citep{Schelling2002,Zhang2015}, or by extrapolating the power spectrum of the energy flux to the zero-frequency limit \citep{Volz2000}. Others have attempted an error analysis of the MD estimate of the GK integral, based on either heuristic or rigorous arguments \citep{Jones2012,Wang_gk2017,Oliveira2017}, but they all require an estimate of an optimal value for the upper limit of integration, which determines a bias in the estimate, and which is in general difficult to obtain. Different classes of systems require different approaches to error analysis, but it is widely believed that all of them always require so long simulation times as to be unaffordable with accurate but expensive AIMD techniques \citep{Carbogno:2017gc}. In order to solve this problem, \cite{Ercole2017} considered it in the light of the statistical theory of stationary time series.

\subsection{Solids and one-component fluids} \label{sec:univariate}
In practice, MD gives access to a discrete sample of the flux process (a \emph{time series}), $J_n = J(n \epsilon)$, $0 \leq n \leq N-1$, where $\epsilon$ is the sampling period of the flux and $N$ the length of the time series, that we assume to be even. As was shown in Sec.~\ref{sec:Einstein}, the Wiener-Khintchine theorem allows one to express the heat conductivity in terms of the zero-frequency value of the power spectrum of the energy-flux (see Eqs.~(\ref{eq:Wiener-Khintchine}-\ref{eq:GK-S0})):
\begin{equation}
\kappa = \frac{\rOmega}{2k_B T^2} S (\omega=0). \label{eq:kappa-S0}
\end{equation}
Let us define the discrete Fourier transform of the flux time series as:
\begin{equation}
  \tilde{J}_{k}=\sum_{n=0}^{N-1} \mathrm{e}^{ 2\pi i\frac{kn}{N}} J_n, \label{eq:Jk}
\end{equation}
for $0 \leq k \leq N-1$.\footnote{Here, the convention for the sign in the exponential of the time-to-frequency Fourier transform is opposite to what adopted in \citep{Ercole2017} and in most of the signal analysis literature, in order to comply with the convention for the space-time Fourier transforms usually adopted in the Physics literature and in Eqs.~\eqref{eq:kontinuity} and \eqref{eq:Fourier-continuity}.} The \emph{sample spectrum} $\hat S_k$, aka \emph{periodogram}, is defined as
\begin{equation}
\hat{S}_{k}=\frac{\epsilon}{N} \left |\tilde{J}_{k} \right |^2, \label{eq:periodogram-def}
\end{equation}
and, for large $N$, it is an unbiased estimator of the power spectrum of the process, as defined in Eq.~\eqref{eq:Wiener-Khintchine}, evaluated at $\omega_k=2\pi\frac{k}{N\epsilon}$, namely: $\langle \hat S_k \rangle = S(\omega_k)$. The reality of the $\hat J$'s implies that $\tilde J_k=\tilde J^*_{N-k}$ and $\hat S_k=\hat S_{N-k}$, so that periodograms are usually reported for $0\leq k\leq \frac{N}{2}$ and their Fourier transforms evaluated as discrete cosine transforms.

The space autocorrelations of conserved currents are usually short-ranged. Therefore, in the thermodynamic limit the corresponding fluxes can be seen as sums of (almost) independent identically distributed stochastic variables, so that, according to the central-limit theorem, their equilibrium distribution is Gaussian. A slight generalization of this argument allows us to conclude that any conserved-flux process is Gaussian as well. The flux time series is in fact a multivariate stochastic variable that, in the thermodynamic limit, results from the sum of (almost) independent variables, thus tending to a multivariate normal deviate. This implies that at equilibrium the real and imaginary parts of the $\tilde J_k$'s defined in Eqs.~\eqref{eq:Jk} are zero-mean normal deviates that, in the large-$N$ limit, are uncorrelated among themselves and have variances proportional to the power spectrum evaluated at $\omega_k$. For $k=0$ or $k=\frac{N}{2}$, $\tilde J_k$ is real and $\sim \mathcal{N}\left (0, \frac{N}{\epsilon}S(\omega_k) \right )$; for $k\notin\left\{ 0,\frac{N}{2}\right\}$, $\mathfrak{Re}\tilde{J}_k$ and $\mathfrak{Im}\tilde{J}_k$ are independent and both  $\sim \mathcal{N}\left (0, \frac{N}{2 \epsilon}S(\omega_k) \right )$, where $\mathcal{N} (\mu,\sigma^2)$ indicates a normal deviate with mean $\mu$ and variance $\sigma^2$. We conclude that in the large-$N$ limit the sample spectrum of the heat-flux time series reads:
\begin{equation}
\hat{S}_{k} = S \left(\omega_k\right) {\xi}_{k}, \label{eq:periodogram-distribution}
\end{equation}
where the $ {\xi}$'s are independent random variables distributed as a $\chi_1^2$ variate for $k=0$ or $k=\frac{N}{2}$ and as one half a $\chi_2^2$ variate, otherwise. Here and in the following $\chi^2_\nu$ indicates the chi-square distribution with $\nu$ degrees of freedom. For the sake of simplicity, we make as though all the ${\xi}$'s were identically distributed, $\xi_k \sim \frac{1}{2} \chi_2^2$ for all values of $k$, thus making an error of order $\mathcal{O}(1/N)$, which vanishes in the long-time limit that is being assumed throughout this section.

In many cases of practical interest, multiple time series are available to estimate the power spectrum of a same process, $\{{^p\!}J_n\}$, $p=1, \cdots \ell$. For instance, in equilibrium MD a same trajectory delivers one independent time series per Cartesian component of the heat flux, all of which are obviously equivalent in isotropic systems. In these cases it is expedient to define a mean sample spectrum by averaging over the $\ell$ different realizations,
\begin{equation}
    \begin{aligned}
      {^{\ell\!}\hat{S}}_{k}& = \frac{\epsilon}{\ell N} \sum_{p=1}^{\ell}  \left |{^p\!}{\tilde J}_{k} \right |^2 \\
      & = S \left(\omega_k\right) {^{\ell\!}{\xi}_{k}},
    \end{aligned} \label{eq:mean-periodogram}
\end{equation}
where the ${^{\ell\!}\xi}$'s are $\chi_{2\ell}^2$ variates, divided by the number of degrees of freedom: $^{\ell\!}\xi_{k}\sim\frac{1}{2\ell}\chi_{2\ell}^{2}$ $\bigl (\text{for } k \notin \{ 0,\frac{N}{2} \}\bigr )$.

\begin{figure}
\centering
\includegraphics[width=8cm]{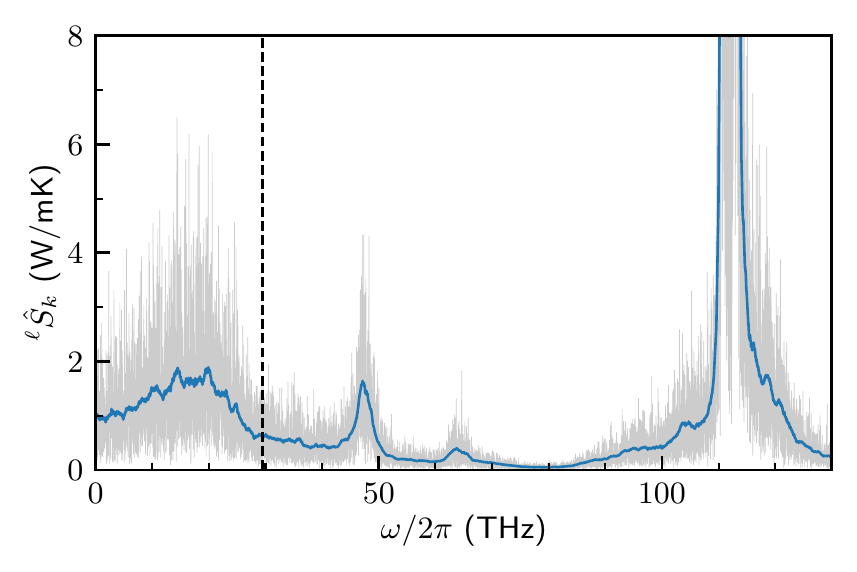}
\caption{Periodogram of a classical flexible model of water obtained from a $100\un{ps}$ MD trajectory. Grey: periodogram obtained directly from Eq.~\eqref{eq:mean-periodogram}, with $\ell=3$. Blue: periodogram filtered with a moving average window of width $1\un{THz}$, useful to reveal the main features of the spectrum (see text). The vertical dashed line delimits the low-frequency region used in the subsequent cepstral analysis.}  \label{fig:water-periodogram}
\end{figure}

Eqs.~\eqref{eq:periodogram-distribution}) and \eqref{eq:mean-periodogram} show that ${^{\ell\!}}{\hat S_0}$ is an unbiased estimator of the zero-frequency value of the power spectrum, $\langle {^{\ell\!}}{\hat S_0} \rangle = S(0)$, and through Eq.~\eqref{eq:kappa-S0}, of the transport coefficients we are after. However, this estimator is not consistent, \emph{i.e.} its variance does not vanish in the large-$N$ limit. This is so because a longer time series increases the number of discrete frequencies at which the power spectrum is sampled, rather than its accuracy at any one of them.

Fig.~\ref{fig:water-periodogram} displays the periodogram of water at ambient conditions, obtained from a $100\un{ps}$ classical MD trajectory, showing the extremely noisy behavior of the periodogram as an estimator of the spectrum. Averaging over the values of the periodogram within a frequency window of given width \citep{MovingAverage} would consistently reduce the statistical noise, but the multiplicative nature of the latter in Eq.~\eqref{eq:periodogram-distribution} makes it difficult to disentangle the noise from the signal and may introduce a bias. In order to cope with this problem, we had better transform the multiplicative noise into an additive one by defining the log-periodogram, $^{{\ell\!}}\hat{L}_{k}$, as:
\begin{equation}
  \begin{aligned}
    ^{{\ell\!}}\hat{L}_{k} &= \log \left (^{\ell\!}\hat{S}_{k} \right ) \\
    &= \log\left(S(\omega_k) \right) + \log\left( ^{\ell\!}{\xi}_k\right ) \\
    &= \log\left(S(\omega_k) \right) + {^{\ell\!}\rLambda} + {^{\ell\!}{\lambda}}_{k},
  \end{aligned} \label{eq:log-PSD}
\end{equation}
where $^{\ell\!}{\lambda}_k = \log\left( {^{\ell\!}{\xi}}_k\right) - {^\ell{\rLambda}}$
are zero-mean identically distributed independent stochastic variables, ${^{\ell\!}\rLambda} = \left\langle \log\left( {^{\ell\!}{\xi}}\right ) \right\rangle = \psi(\ell)-\log(\ell)$, and $\psi(z)$ and is the digamma function \citep{PolyGamma}. The variance of the $^{\ell\!}\lambda$ variables is $\sigma_{\ell}^{2} =\psi'(\ell)$, where $\psi'(z)$ is the tri-gamma function \citep{PolyGamma}.

\begin{figure}
\centering
\includegraphics[width=8cm]{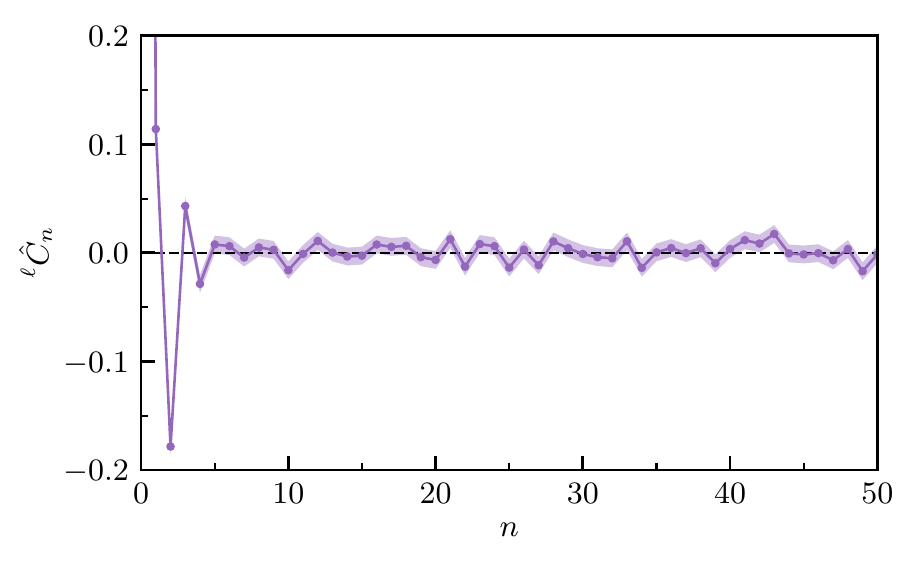}
\caption{Cepstral coefficients of water computed analyzing the low-frequency region of the periodogram (see Fig.~\ref{fig:water-periodogram}), defined in Eq.~\eqref{eq:sample-cepstrum}. }  \label{fig:water-cepstrum}
\end{figure}

Whenever the number of (inverse) Fourier components of the logarithm of the power spectrum is much smaller than the length of the time series, applying a low-pass filter to Eq.~\eqref{eq:log-PSD} would result in a reduction of the power of the noise, without affecting the signal. In order to exploit this idea, we define the ``\emph{cepstrum}'' of the time series as the inverse Fourier transform of its sample log-spectrum \citep{Childers1977}:
\begin{equation}
  ^{\ell\!} \hat C_{n} = \frac{1}{N}\sum_{k=0}^{N-1} {^{\ell\!} \hat L_{k}} \mathrm{e}^{-2\pi i\frac{kn}{N}}. \label{eq:sample-cepstrum}
\end{equation}
A generalized central-limit theorem for Fourier transforms of stationary time series ensures that, in the large-$N$ limit, these coefficients are a set of independent (almost) identically distributed zero-mean normal deviates \citep{Anderson1994,Peligrad2010}. It follows that:
\begin{equation}
  \begin{aligned}
    ^{\ell\!} \hat  C_{n} &= \lambda_{\ell} \delta_{n0} + C_{n} +  {^{{\ell\!}}{\mu}}_{n}, \\
    C_{n} &= \frac{1}{N}\sum_{k=0}^{N-1} \log\bigl (S(\omega_k) \bigr ) \mathrm{e}^{-2\pi i\frac{kn}{N}},
  \end{aligned} \label{eq:cepstrogram}
\end{equation}
where $^{{\ell\!}}{\mu}_{n}$ are independent zero-mean \emph{normal} deviates with variances $\left\langle {^{{\ell\!}}{\mu}_{n}^2}  \right\rangle$ $=\frac{1}{N}\sigma_\ell$ for $n\notin\left\{ 0,\frac{N}{2}\right\}$ and $\left\langle ^{{\ell\!}}{\mu}_{n}^{2}\right\rangle =\frac{2}{N}\sigma_{\ell}^{2}$
otherwise.
Fig.~\ref{fig:water-cepstrum} displays the cepstral coefficients of the low-frequency region of the spectrum of water (marked in Fig.~\ref{fig:water-periodogram}), showing that only the first few coefficients are substantially different from zero.

Let us indicate by $P^*$ the smallest integer such that $C_n \approx 0$ for $P^* \le n \le N-P^*$. By limiting the Fourier transform of the sample cepstrum, Eq.~\eqref{eq:sample-cepstrum}, to $P^*$ coefficients, we obtain an efficient estimator of the zero-frequency component of the log-spectrum as:
\begin{equation}
  \begin{aligned}
    ^{{\ell\!}}\hat{L}_{0}^{*} & = {^{\ell\!}\hat{C}}_{0}+2\sum_{n=1}^{P^{*}-1}{^{{\ell\!}}\hat{C}}_{n} \\
    & = {^{\ell\!}\rLambda} + \log(S_0) + {^{{\ell\!}} {\mu}_{0}}+ 2 \sum_{n=1}^{P^*-1} {^{\ell\!} {\mu}_{n}}.
  \end{aligned} \label{eq:L0*}
\end{equation}
Inspection of Eq.~\eqref{eq:L0*} shows that $^{\ell\!}\hat{L}_{0}^{*}$ is a normal estimator whose expectation and variance are:
\begin{align}
	\langle {^{{\ell\!}}\hat{L}_{0}^{*}}\rangle &= \log(S_{0}) + {^{\ell\!}\rLambda}, \label{eq:L*} \\
	\sigma_\ell^{*}(P^{*},N)^{2} &=\sigma_{\ell}^{2}\frac{4P^{*}-2}{N}. \label{eq:sigma*}
\end{align}
Using Eq.~\eqref{eq:kappa-S0}, we see that the logarithm of the conductivity can be estimated from the cepstral coefficients of the flux time series through Eqs.~(\ref{eq:L0*}-\ref{eq:sigma*}), and that the resulting estimator is always normal with a variance that depends on the specifc system \emph{only} through the number of these coefficients, $P^*$. Notice that the absolute error on the logarithm of the conductivity directly and nicely yields the relative error on the conductivity itself.

\begin{figure}
    \centering
    \begin{subfigure}[tb]{1\textwidth}
    	\centering
        \includegraphics[width=8cm]{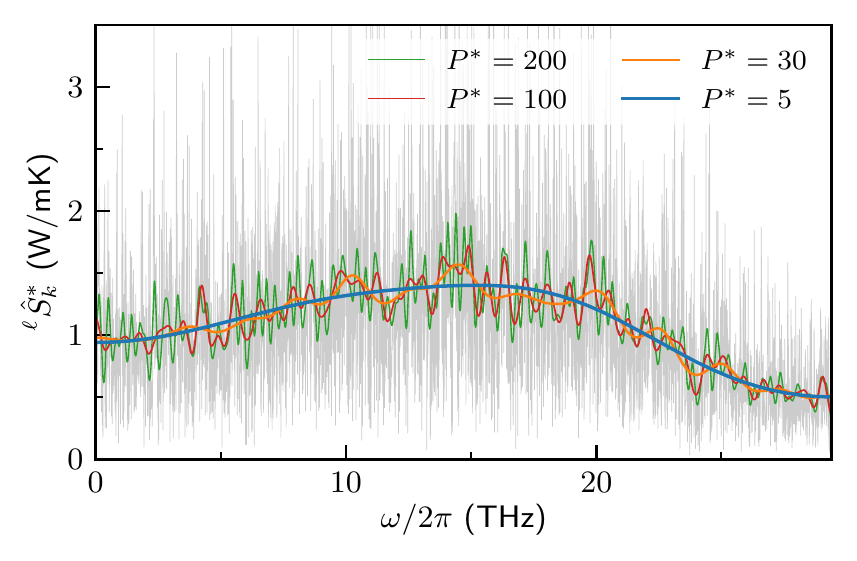}
        \caption{}
        \label{fig:water-filtered-psd}
    \end{subfigure}

    \begin{subfigure}[tb]{1\textwidth}
    	\centering
        \includegraphics[width=8cm]{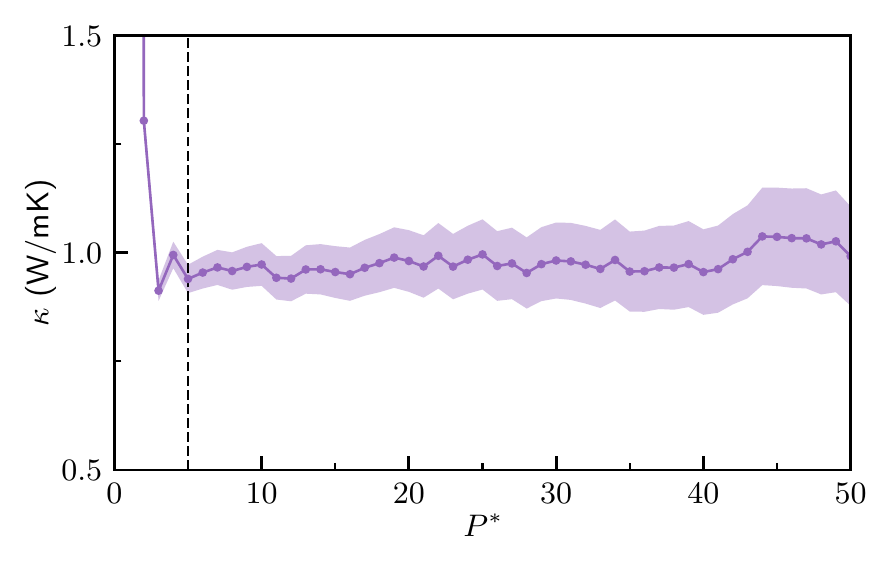}
        \caption{}
        \label{fig:water-fkappa-Pstar}
    \end{subfigure}
    \caption{
    (a) Filtered low-frequency region of the power spectrum of water obtained by limiting the number of cepstral coefficients to various values of $P^*$, Eq.~\eqref{eq:filtered-psd}. $P^*=7$ is the cutoff value suggested by the Akaike's information criterion, Eq.~\eqref{eq:AIC-P}. Grey: the unfiltered periodogram obtained from Eq.~\eqref{eq:periodogram-def}.
    (b) Thermal conductivity of water estimated from Eqs.~(\ref{eq:L0*}-\ref{eq:sigma*}) as a function of the cutoff, $P^*$. The colored bands indicate one standard deviation as estimated from theory. The vertical dashed line indicates the value suggested by the Akaike's information criterion, Eq.~\eqref{eq:AIC-P}.
    }
\end{figure}

The efficacy of this approach obviously depends on our ability to estimate the number of coefficients necessary to keep the bias introduced by the truncation to a value smaller than the statistical error, while maintaining the magnitude of the latter at a prescribed acceptable level. \cite{Ercole2017} proposed to estimate $P^*$ using the Akaike's information criterion (\cite{Akaike1974}), but other more advanced \emph{model selection} approaches \citep{Claeskens2008} may be more effective. This method consists in choosing $P^*$ as the one that minimizes the function:
\begin{equation}
\mathrm{AIC}(P)=\frac{N}{\sigma_\ell^{2}}\sum_{n=P}^\frac{N}{2} \hat{C}_{n}^{2}+2P. \label{eq:AIC-P}
\end{equation}
In Fig.~\ref{fig:water-filtered-psd} we report the low-frequency region of the spectrum of water obtained by limiting the number of cepstral coefficients to $P^*$:
\begin{equation}
^\ell\hat{S}_k^* = \exp\left[ 2\sum_{n=1}^{P^*-1} {}^\ell\hat{C}_n \mathrm{e}^{2\pi i \frac{k n}{N}} + {}^\ell\hat{C}_0 - {}^\ell\rLambda\right], \label{eq:filtered-psd}
\end{equation}
thus showing the filtering effect of this choice.
Finally, Fig.~\ref{fig:water-fkappa-Pstar} shows the value of thermal conductivity of water obtained through Eqs.~(\ref{eq:L0*}-\ref{eq:sigma*}).

\subsection{Multi-component fluids}
In Sec.~\ref{sec:multi-component} we have seen that in a fluid made of $Q$ atomic species there are in general $Q$ macroscopic fluxes interacting with each other through Onsager's phenomenological equations, Eq.~\eqref{eq:onsager}, not counting the different Cartesian components that do not interact amongst themselves because of space isotropy. A MD simulation thus samples $Q$ stochastic processes, one for each interacting flux, that we suppose to be stationary. These processes can be thought of as different components of a same multivariate process \citep{Bertossa2018}. As in Sec.~\ref{sec:univariate}, for the sake of generality we suppose to have $\ell$ independent samples of such a process, described by a multivariate time series of length $N$: $\{ ^{p\!}{J}^i_n \}$; $p=1,\dots \ell$; $i=1,\dots Q$; $n=0,\dots N-1$. Stationarity implies that $\langle {J}^i_n\rangle $ does not depend on $n$ and that $\langle {J}^i_n {J}^j_m \rangle$ only depends on $n-m$. We will further assume that $\langle {J}^i_n\rangle =0 $ and that $\langle {J}^i_n {J}^j_0 \rangle$ is an even function of $n$, which is the case when ${J}^i$ and ${J}^j$ have the same signature under time-reversal. By combining Eq.~\eqref{eq:multi_kappa} with Eq.~\eqref{eq:GK-S0}, we see that in order to evaluate the thermal conductivity in the multi-component case we need an efficient estimator for $\left ( S^{-1}_0\right )^{11}$, where $S^{kl}_0=S^{kl}(\omega=0)$ is the zero-frequency cross-spectrum of the relevant fluxes, ordered in  such a way that the energy one is the first.

Similarly to the one-component case, we define a mean sample cross-spectrum (or \emph{cross-periodogram}) as
\begin{equation}
 ^{(\ell Q)\!}\hat{S}_k^{ij} = \frac{1}{\ell} \sum_{p=1}^{\ell} \frac{\epsilon}{N} \left({}^{p\!}\tilde{J}_k^i\right)^* {}^{p\!}\tilde{J}_k^j .
\end{equation}
By discretizing Eq.~\eqref{eq:Sij(omega)} we see that $^{(\ell Q)\!}\hat{S}_k^{ij}$ is an unbiased estimator of the cross-spectrum, $\left \langle {}^{(\ell Q)\!}\hat{S}_k^{ij} \right \rangle = S^{ij}\left (\omega_k= \frac{2\pi k }{N\epsilon}\right )$. As it was the case for univariate processes, in the large-$N$ limit the real and imaginary parts of $\tilde J^i_k$ are normal deviates that are uncorrelated for $k\ne k'$. We conclude that the cross-periodogram is a random matrix distributed as a complex Wishart deviate \citep{Goodman1963a,Goodman1963b}:
\begin{equation}
  {}^{(\ell Q)\!}\hat{S}_k \sim \mathcal{CW}_Q \left(S(\omega_k), \ell\right). \label{eq:ComplexWishart}
\end{equation}
The notation $\mathcal{CW}_Q \left(S, \ell \right)$ in Eq.~\eqref{eq:ComplexWishart} indicates the distribution of the $Q\times Q$ Hermitian matrix
${}^{(\ell Q)\!}\hat{S}^{ij} = \frac{1}{\ell}\sum_{p=1}^\ell  {}^{p\!}{X}^i \, {}^{p\!}{X}^{j*}$,
where $\{ {}^{p\!}{X}^i \}$ ($p=1,\cdots\ell$, $i=1, \cdots Q$) are $\ell$ samples of an $Q$-dimensional zero-mean normal variate whose covariance is $S^{ij} = \langle X^i X^{j*} \rangle $.

\begin{figure}
\centering
\includegraphics[width=8cm]{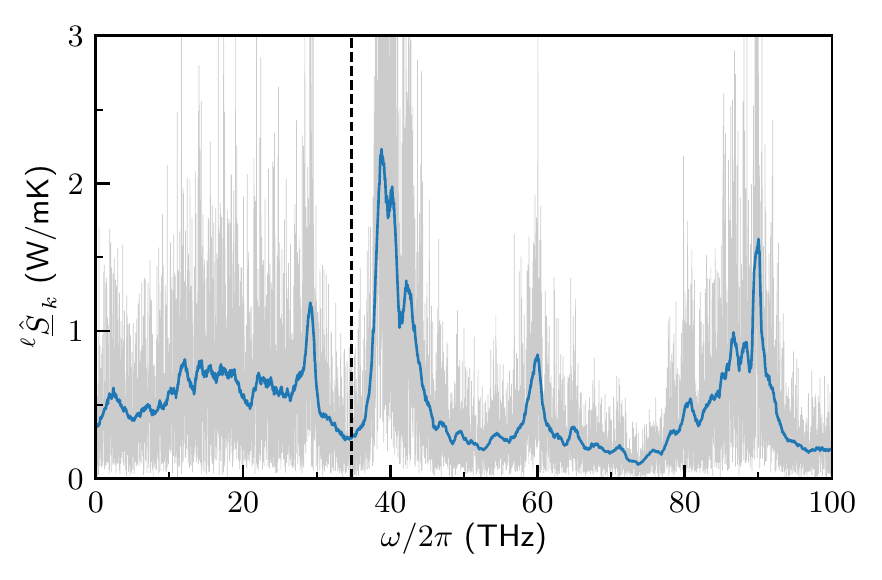}
\caption{Multi-component power spectrum, as defined in Eq.~\eqref{eq:mean-multi-periodogram}, for a classical flexible model of a solution of water and ethanol $50\un{mol}\%$, obtained from a $100\un{ps}$ trajectory. Grey: $^{\ell}\hat{\underline{S}}_{\,k}$ obtained directly from Eq.~\eqref{eq:mean-multi-periodogram}, with $\ell=3$ and $Q=2$. Blue: $^{\ell}\hat{\underline{S}}_{\,k}$ filtered with a moving average window of width $1\un{THz}$ in order to reveal its main features. The vertical dashed line delimits the low-frequency region used in the subsequent cepstral analysis. Reproduced from \cite{Bertossa2018}.}  \label{fig:grappa-periodogram}
\end{figure}

Similarly to the real case, a Bartlett decomposition \citep{kshirsagar1959} holds for complex Wishart matrices \citep{Nagar2011}, reading:
\begin{equation}
{}^{(\ell Q)\!}\hat{S} = \frac{1}{\ell} \mathcal{S} R R^\top \mathcal{S}^{\dagger},  \label{eq:S_cholesky}
\end{equation}
where ``$\top$'' and ``$\dagger$'' indicate the transpose and the adjoint of a real and complex matrix, respectively; $\mathcal{S}$ is the Cholesky factor of the covariance matrix, $S= \mathcal{S} \mathcal{S}^{\dagger}$, and $R$ is a real lower triangular random matrix of the form
\begin{equation}
 R =
 \begin{pmatrix}
 c_1 & 0 & 0 & \cdots & 0\\
 n_{21} &  c_2 &0 & \cdots& 0 \\
 n_{31} &  n_{32} &  c_3 & \cdots & 0\\
\vdots & \vdots & \vdots &\ddots & \vdots \\
 n_{\smallQ1} & n_{\smallQ2} & n_{\smallQ3} &\cdots & c_\smallQ
 \end{pmatrix},
\end{equation}
where $c^2_i \sim \chi^2_{2(\ell-i+1)}$ and $ n_{ij}\sim \mathcal{N}(0,1)$. We stress that $R$ is independent of the specific covariance matrix, and only depends upon $\ell$ and $Q$. In particular it is independent of the ordering of the fluxes $J^i$. By expressing the $QQ$ matrix element of the inverse of $^{(\ell Q)}\hat{S}$ in Eq.~\eqref{eq:S_cholesky} as the ratio between the corresponding minor and the full determinant, and using some obvious properties of the determinants and of triangular matrices, we find that:
\begin{equation}
\frac{\ell}{\left({}^{(\ell Q)}\hat{S}_k^{-1}\right)^{\smallQ\smallQ}} = \frac{1}{\left(S_k^{-1}\right)^{\smallQ\smallQ}} c^2_\smallQ, \label{eq:S-1_choleskied}
\end{equation}
As the ordering of the fluxes is arbitrary, a similar relation holds for all the diagonal elements of the inverse of the cross-periodogram. We conclude that the generalization of Eq.~\eqref{eq:mean-periodogram} for the multi-component case is:
\begin{equation}
   ^{\ell}\hat{\underline{S}}_{\,k}\equiv\frac{\ell}{2(\ell-Q+1)}\frac{1}{\left( {}^{(\ell Q)}\hat{S}_k^{-1} \right)^{\smallone\smallone}} = \frac{1}{\left(S_k^{-1}\right)^{\smallone\smallone}} \, \xi_k, \label{eq:mean-multi-periodogram}
\end{equation}
where $\xi_k$ are independent random (with respect to $k$) random variables, distributed as
\begin{equation}
  \xi_k \sim
  \begin{cases}
    \frac{1}{\ell-Q+1} \,\chi^2_{\ell-Q+1}  \qquad & \mathrm{for} \; k \in \{0 , \frac{N}{2}\}, \\
 \\
 \frac{1}{2(\ell-Q+1)} \, \chi^2_{2(\ell-Q+1)} \qquad & \mathrm{otherwise}.
\end{cases}
\end{equation}
Starting from here we can apply the cepstral analysis as in the one-component case. The only difference is the number of degrees of freedom of the $\chi^2$ distribution, that becomes $2(\ell -Q+1)$, and a different factor in front of the result. Fig.~\ref{fig:grappa-periodogram} shows an example of multi-component power spectrum for a solution of water and ethanol.

\begin{figure}
    \centering
    \begin{subfigure}[tb]{1\textwidth}
    	\centering
        \includegraphics[width=8cm]{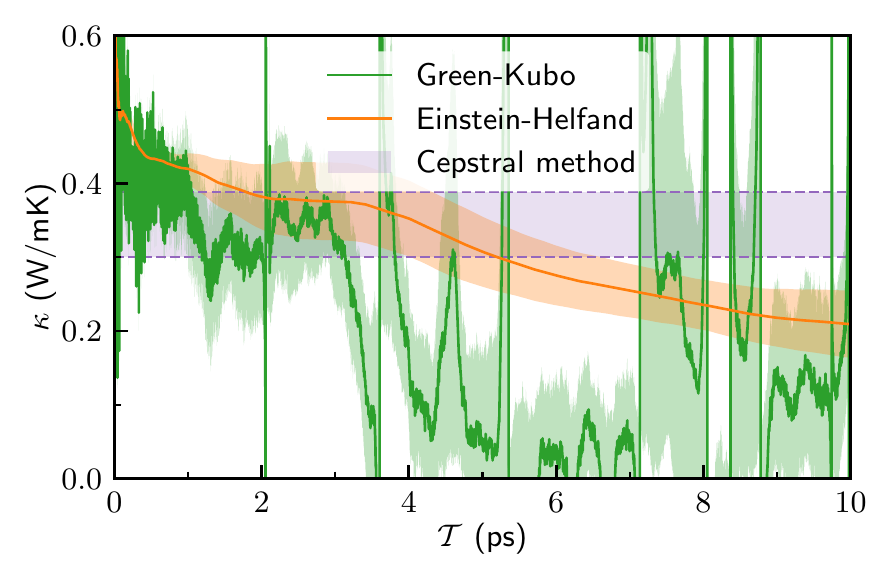}
        \caption{}
    \end{subfigure}

    \begin{subfigure}[tb]{1\textwidth}
    	\centering
        \includegraphics[width=8cm]{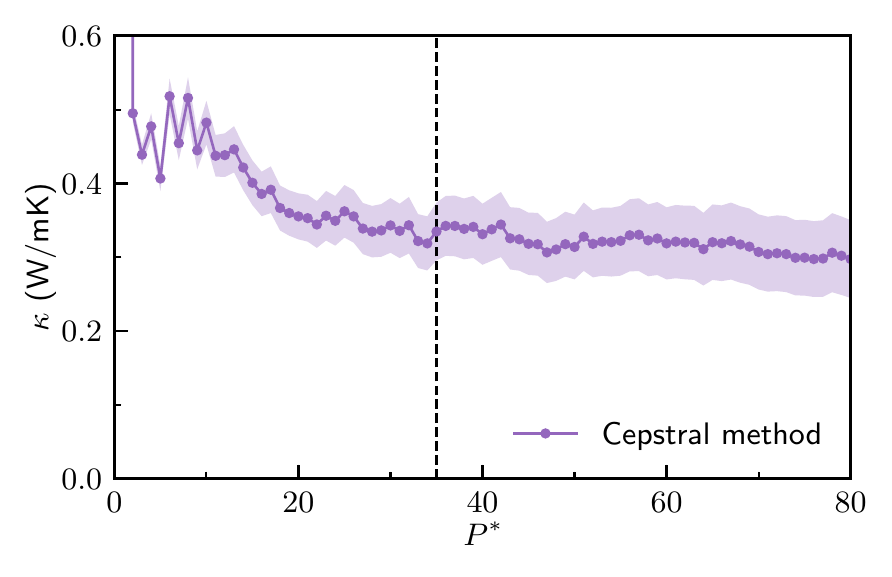}
        \caption{}
    \end{subfigure}
	\caption{Convergence of the multi-component thermal conductivity estimator $\kappa$ using the direct time-integration approach and the cepstral method, for a classical flexible model of a solution of water and ethanol $50\un{mol}\%$, obtained from a $100\un{ps}$ trajectory.
(a) Direct time-integration approach in its Green-Kubo (green, as obtained from the matrix $L^{ij}(\mathcal{T})\propto\int_0^\mathcal{T} \left\langle J^i(t) J^j(0) \right\rangle dt $) and Einstein-Helfand (orange -- obtained from the  matrix $\left (L^{ij}\right )'(\mathcal{T}) \propto  \int_0^\mathcal{T}\left(1-\frac{t}{\mathcal{T}}\right) \left \langle J^i(t) J^j(0) \right \rangle dt$) formulations. The horizontal purple band indicates the value obtained by the cepstral method.
(b) Estimate of $\kappa$ with the cepstral method as a function of the number of cepstral coefficients, $P^*$, see Eqs.~(\ref{eq:L0*}-\ref{eq:sigma*}). The dashed vertical line indicates the value of $P^*$ selected by the AIC, Eq.~\eqref{eq:AIC-P}. Reproduced from \cite{Bertossa2018}.
}  \label{fig:twoCompConvergence}
\end{figure}

The method discussed so far shows a fundamental advantage with respect to a na\"ive implementation of direct time-integration approach.
Fig.~\ref{fig:twoCompConvergence} shows the two-component conductivity $\kappa$, obtained via Eq.~\eqref{eq:two-comp-kappa}, in the case of a water-ethanol solution, as a function of the upper time-integration limit $\mathcal{T}$ \citep{Bertossa2018}. Both the Green-Kubo and the Einstein-Helfand definitions of the finite-time expression of Onsager's coefficients (see Eq.~\eqref{eq:Einstein-Helfand}) are displayed.
Due to thermal fluctuations, the integral of the correlation function becomes a random walk as soon as the latter vanishes, eventually assuming any value. Therefore, there will be a set of times (see Fig.~\ref{fig:twoCompConvergence}) where the term $L^{\smallQ\smallQ}$ at the denominator in Eq.~\eqref{eq:two-comp-kappa} vanishes, leading to divergences in the evaluation of $\kappa$; an issue not affecting the one-component case. Hence, in such a formulation of the multi-component case, the mean value of the thermal conductivity estimator \textit{in the time domain} does not exist. On the contrary, the multi-component frequency-domain approach presented in this section, and built on sound statistical basis, provides a well defined expression for the estimator of $\kappa$ and its statistical error.

\subsection{Data analysis work-flow}
We summarize the steps leading to the estimation of thermal conductivity by the \textit{cepstral analysis} method, in order to highlight the simplicity of its practical implementation.
\begin{enumerate}
\item From a MD simulation compute the heat flux time series $J_n^1$ and the independent particle fluxes $J_n^q$, $q=2,\dots,Q$.
\item Compute the discrete Fourier transform of the fluxes, $\tilde{J}^{\small i}_k$, and the element $1/(\hat{S}^{-1})^{\smallone\smallone}$. In practice, only a selected low-frequency region shall be used (see \cite{Ercole2017} for a detailed discussion).\footnote{To lighten the notation, we drop the left superscripts of the variables in this subsection.}
\item Calculate $\log\left[1/(\hat{S}^{-1})^{\smallone\smallone}\right]$.
\item Compute the inverse discrete Fourier transform of the result to obtain the cepstral coefficients $\hat{C}_n$.
\item Apply the Akaike Information Criterion, Eq.~\eqref{eq:AIC-P}, to estimate the number of cepstral coefficients to retain, $P^*$.
\item Finally apply Eq.~\eqref{eq:L0*} to obtain $\hat{L}_0^*$, and evaluate the thermal conductivity as
\begin{equation}
\kappa = \frac{\rOmega}{2k_B T^2} \exp\left[\hat{L}_0^* - \psi(\ell - Q+1) + \log(\ell -Q+1) \right],
\end{equation}
and its statistical error as
\begin{equation}
\frac{\rDelta\kappa}{\kappa} = \sqrt{\psi'(\ell -Q+1) \frac{4P^{*}-2}{N}}.
\end{equation}
\end{enumerate}

%%%%%%%%%%%%%%%%%%%%%%%%%%%%%%%%%%%%%%%%%%%%%%%%%%%%%%%%%%%%%%%%%%%%%%%%%%%%%%%%%%%%
\section{A few representative results}
Calculations of the thermal conductivity based on the Green-Kubo formalism combined with first-principles molecular dynamics are quite recent. The first benchmarks from \cite{Marcolongo2016} have been performed on liquid Argon and heavy water at ambient conditions, as reported below.

\subsection{A benchmark on a model mono-atomic fluid}
As a first test, liquid Argon was simulated by \cite{Marcolongo2016} using a local LDA functional neglecting dispersion forces. The resulting fictitious system, dubbed LDA-Argon, is a hard-core weakly interacting fluid whose dynamics is expected to be easily mimicked by a simple two-body potential, which can be engineered by standard force-matching techniques. This observation allows one to effectively test the ideas developed in Sec.~\ref{sec:DFT}: even if the energy density in LDA-Argon and in its fitted classical counterpart will likely be different, the resulting thermal conductivity is expected to coincide within the quality of the classical fit.

\begin{figure}
\begin{subfigure}{.5\textwidth}
  \centering
  \includegraphics[scale=0.66]{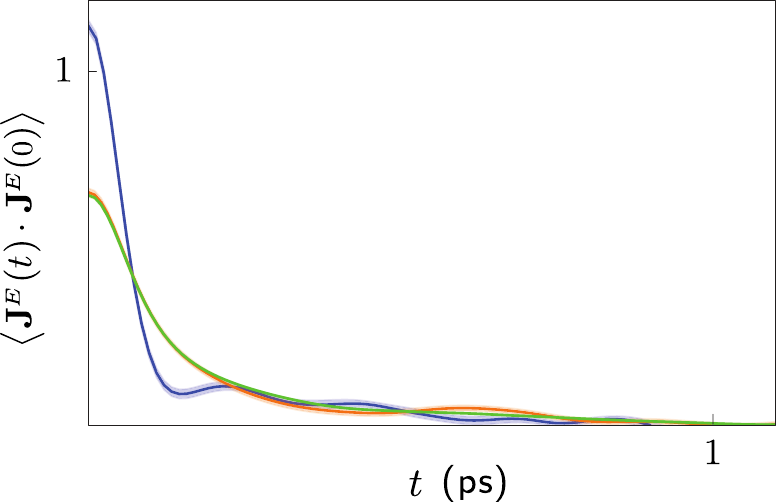}
  \caption{}
\end{subfigure}
\begin{subfigure}{.45\textwidth}
  \centering
  \includegraphics[scale=0.66]{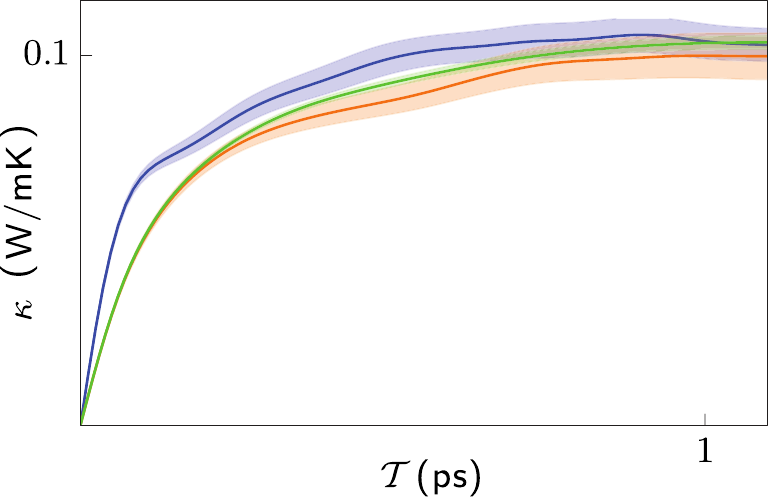}
  \caption{}
\end{subfigure}
\caption{Benchmark of LDA-Argon, reproduced from \cite{Marcolongo2016}.
(a) The heat-flux autocorrelation function.
(b) The GK integral as a function of the upper limit of integration. Color coding common to the two panels. Blue: \emph{ab initio} molecular dynamics ($100\un{ps}$). Orange: classical molecular dynamics ($100\un{ps}$). Green: classical molecular dynamics ($1000\un{ps}$)}
\label{fig:argon-dft}
\end{figure}

Simulations were performed in a cubic supercell of 108 atoms with an edge of $17.5\un{\angstrom}$, corresponding to a density of $1.34\un{g\,cm^{-3}}$. Trajectories were sampled in the NVE ensemble for $100\un{ps}$ and the classical model was fitted with a pair potential of the form $V(r)=P_2(r)\mathrm{e}^{-\alpha r}$, $P_2(r)$ being a second order polynomial. In Fig.~\ref{fig:argon-dft} we show the resulting autocorrelation functions at a representative temperature of $400\un{K}$. The DFT autocorrelation function shows a more structured behavior than that of the classical potential. Nevertheless, when considering the long time limit of the Green-Kubo integral, the thermal conductivities indeed coincide within statistical uncertainty, as predicted by theory.

\subsection{Heavy water at ambient conditions}  \label{sec:Results-water}
\cite{Marcolongo2016} also computed the thermal conductivity of heavy water at ambient conditions. Simulations were performed using the PBE exchange correlation energy functional \citep{Perdew1996}, which is known to predict qualitatively the self-diffusion coefficient of water at ambient conditions only for a simulation temperature of about $400\un{K}$ \citep{Sit-Marzari}: the latter condition was imposed accordingly. A system of 64 heavy-water molecules was sampled for 90 ps in the NVE ensemble using a cubic cell corresponding to the experimental density of $1.11\un{g\,cm^{-3}}$.

The \emph{ab initio} treatment of heat transport in molecular fluids requires some care in order to eliminate non-diffusive components of the energy flux that, while not contributing to the conductivity, do increase the noise of the flux time series to a level that may compromise its analysis. To see where the problem comes from, let us split the potential energy of the system into the sum of non-interacting atomic energies plus an interaction energy, as: $V(\{\mathbf{R}_n\}) = \sum_n\epsilon^\circ_n + V_{int}(\{\mathbf{R}_n\})$, where $\epsilon_n^\circ$ is the energy of the $n$-th atom when it is isolated from the rest. In classical simulations the energy of isolated atoms never enters the description of the system, and the $\epsilon^\circ_n$'s can be simply set to zero. In quantum simulations, instead, atomic and interaction energies enter on a same footing and the former give a large and fluctuating contribution to the total energy flux, $\mathbf{J}^\circ =\sum_X\epsilon^\circ_X \mathbf{J}^X$, where $ \mathbf{J}^X $ is the flux defined in Eq. \eqref{eq:JX}. In a monoatomic fluid $ \mathbf{J}^X $ is constant because of momentum conservation and it is actually equal to zero in the center-of-mass reference frame. In molecular fluids the $ \mathbf{J}^X $ do not vanish but, as we have seen in Sec. \ref{sec:MolecularFluids}, they are non-diffusive and hence do not contribute to the heat conductivity, while adding considerable noise to the energy-flux time series. In order to remove them, instead of estimating $\mathbf{J}^\circ$ from the non-interacting atomic energies, we prefer to implement a \emph{decorrelation technique}, as described below.

Current decorrelation builds on a general inequality whose proof can be found in \cite{Marcolongo2016}.  Let $\mathbf{J}^1$ and $\mathbf{J}^2$ be two macroscopic fluxes and $\mathbf{J}^{12}=\mathbf{J}^1+\mathbf{J}^2$ their sum. The corresponding conductivities $\kappa^1, \kappa^2$, and $\kappa^{12}$ then satisfy $|\kappa^{12}-\kappa^1-\kappa^2| \le 2 \sqrt{\kappa^1 \kappa^2}$. As a consequence, when $\kappa^2$ vanishes, $\kappa^{12}$ coincides with $\kappa^1$. Let us now suppose that a set of fluxes $\lbrace\mathbf{Y}^u\rbrace$, $u=1,\dots U$ is known to exhibit a non-diffusive behavior. The above argument shows that the auxiliary flux defined as
\begin{equation}
 \mathbf{J}' \equiv \mathbf{J} - \sum_w \lambda^w \mathbf{Y}^w,
\end{equation}
will yield the same thermal conductivity as $\mathbf{J}$.
Optimal values of the $\lbrace\lambda^u\rbrace$ coefficients can then be determined by imposing that the new time series $\mathbf{J}'$ is uncorrelated with respect to the non-diffusive ones, \emph{i.e.}:
\begin{equation}
\langle \mathbf{J} \mathbf{Y}^u \rangle-\sum_{w} \lambda^{w} \langle \mathbf{Y}^{w} \mathbf{Y}^u \rangle = 0, \quad u=1,\dots U.
\end{equation}
This procedure is particularly useful when the $\mathbf{Y}^u$ fluxes give a slowly converging contribution to the Green-Kubo integral, which is thus difficult to evaluate numerically.

\begin{figure}
    \centering
    \begin{subfigure}[tb]{1\textwidth}
    	\centering
        \includegraphics[scale=1.0]{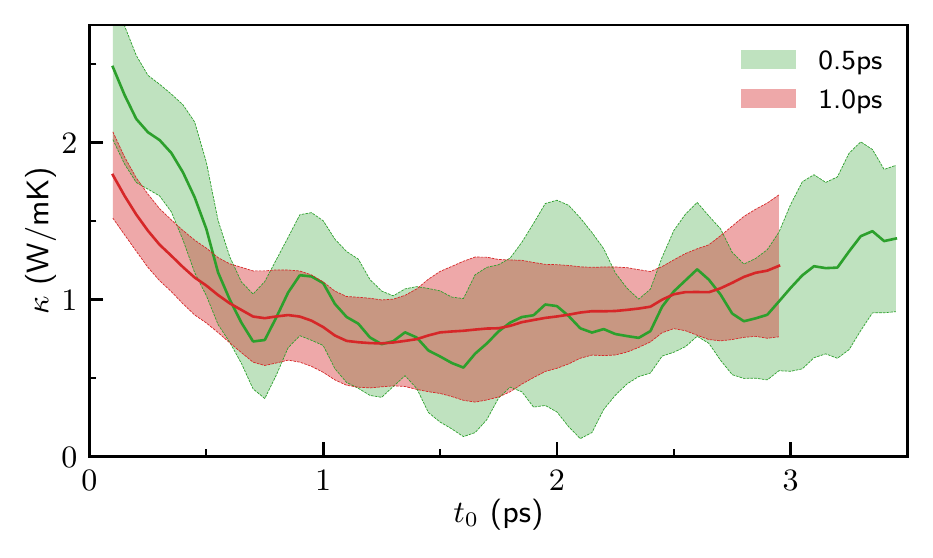}
        \caption{}
        \label{fig:disp}
    \end{subfigure}

    \begin{subfigure}[tb]{1\textwidth}
    	\centering
        \includegraphics[scale=1.0]{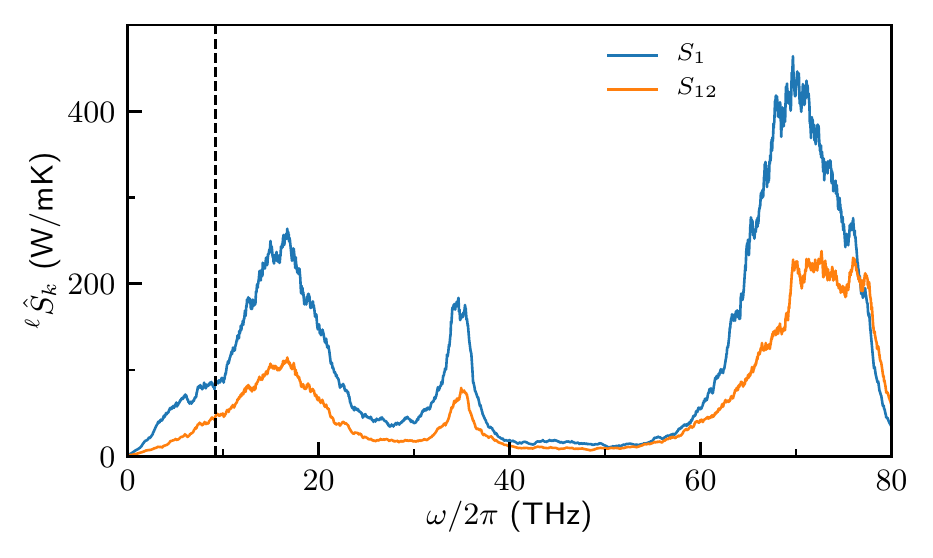}
        \caption{}
        \label{fig:water_psd}
    \end{subfigure}
    \caption{(a) Value of thermal conductivity of heavy water obtained from a linear fitting of the energy displacement of
    $\left ( \mathbf{J}^1 \right )'$, $\mathcal{D}(\tau)=\int_0^\tau \left ( \mathbf{J}^1 \right )'(t)dt$. The two curves refer to different window widths used for the linear fit, of length $0.5$ and $1.0\un{ps}$. The abscissa corresponds to the origin of the fitting window. (b) Periodogram of the $\left ( \mathbf{J}^1 \right )'$ and $\left ( \mathbf{J}^{12} \right ) '$ currents, filtered with a moving average window in order to reveal the prominent features. The vertical dashed line delimits the low-frequency region used for cepstral analysis.}
\end{figure}

The decorrelation technique has been applied to heavy water considering two non-diffusive number fluxes: $\mathbf{Y}^1 = \mathbf{J}^H + \mathbf{J}^O$,  \emph{i.e.} the sum of hydrogen and oxygen average velocities,\footnote{Note that the two time series $\mathbf{J}^H$ and $\mathbf{J}^O$ are trivially related, because of momentum conservation. Therefore $\mathbf{J}^H$, $\mathbf{J}^O$, or $\mathbf{J}^H+\mathbf{J}^O$ would all be equivalent choices.} and $\mathbf{Y}^2 = \mathbf{J}^{el}$, the adiabatic electronic current. The latter is defined, following the same notation of Sec. \ref{sec:DFT}, as:
\begin{equation}
 \mathbf J^{el} = \frac{2}{\rOmega}\mathfrak{Re} \sum_{v} \langle\bm{\bar \phi}_v^c |\dot \phi_v^c \rangle ,
\end{equation}
as can be derived from the continuity equation for the density: $\nabla\cdot \bm j^{el}(\bm r,t) =- \dot{n}^{el}(\bm r,t)$. In insulators $\mathbf J^{el}$ is non-diffusive and can thus be used to decorrelate the heat current. In the original paper the thermal conductivity was evaluated from the slope of the energy displacement $\mathcal{D}(\tau)=\int_0^\tau \mathbf{J}(t)dt$ (see Eq.~\eqref{eq:Einstein-Helfand}) and the corresponding error obtained from a standard block analysis, resulting in a non optimal estimate of both (see Fig.~\ref{fig:disp}). The same data have been re-analyzed with the novel cepstral technique presented in Sec.~\ref{sec:data-analysis}. We denote by $\left( \mathbf{J}^1\right)'$ and $\left (\mathbf{J}^{12} \right )'$ the currents decorrelated with respect to $\mathbf{Y}^1$ alone, and with respect to both $\mathbf{Y}^1$ and $\mathbf{Y}^2$, respectively. The power spectra of the two currents are plotted in Fig.~\ref{fig:water_psd}. Only the low-frequency region of the spectra (up to $\sim 9.0 \un{THz} $) was used for the cepstral analysis (see \cite{Ercole2017} for the technical details).  The minimization of Eq.~\eqref{eq:AIC-P} suggests a cutoff $P^*_1=17$ and $P^*_{12}=15$, but a value $1.5$ times larger was actually been used to reduce the bias possibly due to the fast variation of the spectrum at frequency close to zero. The resulting thermal conductivities are $\kappa^1 =  0.80 \pm 0.12 \un{W/mK}$ for the $\left( \mathbf{J}^1 \right)'$ flux, and $\kappa^{12} = 0.93 \pm 0.14 \un{W/mK}$ for $\left ( \mathbf{J}^{12} \right )'$, compatible with each other. By comparison, experiments give a value $\kappa\approx 0.6\un{W/mK}$ \citep{Matsunaga,Ramires}.
In this case, data analysis would not have yielded any meaningful results failing a proper decorrelation of the heat flux time series.

%%%%%%%%%%%%%%%%%%%%%%%%%%%%%%%%%%%%%%%%%%%%%%%%%%%%%%%%%%%%%%%%%%%%%%%%%%%%%%%%%%%%
\section{Outlook}
We believe that the ideas presented in this chapter will pave the way to new developments and applications in the field of heat transport, particularly for strongly anharmonic and/or disordered systems, where approaches based on the Boltzmann transport equation do not apply or are bound to fail. The general concept of gauge invariance of heat conductivity will likely apply to other transport properties as well, such as ionic conduction, viscosity, and many others, and/or simulation methodologies, such as those based on a neural-network representation of interatomic potentials, which hold the promise of a strong and long-lasting impact on molecular simulations. The applicability of this concept would not be as broad if not assisted by the powerful data-analysis methods which have also been described in this chapter. Here again, we believe that there is ample room for improvement, leveraging more general (possibly non-Fourier) representations of the log-spectrum of the currents to be analyzed, and more advanced statistical-inference techniques to estimate the parameters of the spectral models resulting from these representations. From the applicative point of view, we expect that these methodological advances will have a strong impact in all those cases where heat, mass, and charge transport occurs in conditions that cannot be adequately described at the atomistic level by simple force fields, such as, \emph{e.g.}, in complex materials, systems at extreme external conditions, such as those occurring in the planetary interiors, complex or reactive fluids, and many others.

%%%%%%%%%%%%%%%%%%%%%%%%%%%%%%%%%%%%%%%%%%%%%%%%%%%%%%%%%%%%%%%%%%%%%%%%%%%%%%%%%%%%
%%%%%%%%%%%%%%%%%%%%%%%%%%%%%%%%%%%%%%%%%%%%%%%%%%%%%%%%%%%%%%%%%%%%%%%%%%%%%%%%%%%%
\begin{acknowledgement}
This work was supported in part by the \textsc{MaX} EU Centre of Excellence, grant no 676598. SB, LE, and FG are grateful to Davide Donadio for insightful discussions all over the Summer of 2017 and beyond.
\end{acknowledgement}

% For bibtex users:
% For references use the `Springer Basic Style'.
\bibliographystyle{spbasic}
\bibliography{bibliography}

\end{document}